\documentclass[apj]{emulateapj}
\usepackage{multirow,amsmath,natbib,longtable,appendix}
\received{August 21, 2014} 
\accepted{October 28, 2014}

\shorttitle{The bulge - black hole scaling relation}
\shortauthors{Scott \& Graham}

\begin{document}

\title{The (black hole)-bulge mass scaling relation at low masses}

\author{Alister W.\ Graham}
\affil{Centre for Astrophysics and Supercomputing, Swinburne University of
  Technology, Hawthorn, VIC, 3122, Australia.}

\and

\author{Nicholas Scott} 
\affil{Sydney Institute for Astronomy, School of Physics, University of
Sydney, NSW 2006, Australia; ARC Centre of Excellence for All-sky Astrophysics.} 

\begin{abstract}

  Several recent papers have reported on the occurrence of active galactic
  nuclei (AGN) containing {\it under-massive} black holes relative to a linear
  scaling relation between black hole mass ($M_{\rm bh}$) and host spheroid
  stellar mass ($M_{\rm sph,*}$).  Dramatic revisions to the $M_{\rm
    bh}$-$M_{\rm sph,*}$ and $M_{\rm bh}$-$L_{\rm sph}$ relations, based on
  samples containing predominantly inactive galaxies, have however recently
  identified a new steeper relation at $M_{\rm bh} \lesssim (2$--$10)\times
  10^8$ M$_\odot$, roughly corresponding to $M_{\rm sph,*} \lesssim
  (0.3$--$1)\times 10^{11}$ M$_\odot$.  We show that this steeper,
  quadratic-like $M_{\rm bh}$--$M_{\rm sph,*}$ relation defined by the
  S\'ersic galaxies, i.e.\ galaxies without partially depleted cores, roughly
  tracks the apparent offset of the AGN having $10^5 \lesssim M_{\rm
    bh}/M_{\odot} \lesssim 0.5\times10^8$.  That is, these AGN are not
  randomly offset with low black hole masses, but also follow a steeper
  (non-linear) relation.  As noted by Busch et al., confirmation or rejection
  of a possible AGN offset from the steeper $M_{\rm bh}$-$M_{\rm sph,*}$
  relation defined by the S\'ersic galaxies will benefit from improved stellar
  mass-to-light ratios for the spheroids hosting these AGN.  Several
  implications for formation theories are noted.  Furthermore, reasons for
  possible under- and over-massive black holes, the potential existence of
  intermediate mass black holes ($<10^5 M_{\odot}$), and the new steep (black
  hole)--(nuclear star cluster) relation, $M_{\rm bh} \propto M_{\rm
    nc}^{2.7\pm0.7}$, are also discussed.

\end{abstract}


\section{Introduction}
\label{sec:intro}

Several years ago, \citet{Graham:2007a,Graham:2008a,Graham:2008b} and \citet{Hu:2008}
reported on galaxies whose black hole masses, $M_{\rm bh}$, appeared
under-massive relative to expectations based on their stellar velocity
dispersion, $\sigma$. This apparent sub-structure in the $M_{\rm
  bh}$--$\sigma$ diagram \citep{FaM:2000,Geb:2000} was due to barred galaxies 
located, on average, 0.3 dex in the $\log M_{\rm bh}$-direction below 
the $M_{\rm bh}$--$\sigma$ relation defined by barless galaxies.  \citet{GandLi:2009}
subsequently revealed that galaxies with active galactic nuclei (AGN)
also display this same general separation in the $M_{\rm bh}$--$\sigma$
diagram, supporting the earlier introduction of barred-, barless-, and
elliptical-galaxy $M_{\rm bh}$--$\sigma$ relations 
\citep[see also][]{Gultekin:2009,Greene:2010}.  It was noted from the start that either
the black hole masses could be low in the barred galaxies, or that an elevated
velocity dispersion may account for their apparent offset in the $M_{\rm
  bh}$--$\sigma$ diagram. 
\citet{Hartmann:2013} \citep[see
  also][]{Brown:2013,Debattista:2013,Monari:2014} have recently used
simulations to demonstrate that the observed offset is an expected result from bar
dynamics, which inflate the measured velocity dispersion by exactly the amount
observed \citep{Graham:2011}.  Given that this can fully account for 
the offsets in the $M_{\rm bh}$--$\sigma$ diagram, it implies that the barred
galaxies do not possess under-massive black holes, and thus should not be
offset in the black hole mass -- spheroid mass ($M_{\rm bh}$--$M_{\rm sph}$)
diagram.

A few recent papers \citep[e.g.][]{Jiang:2011a,Jiang:2013,Mathur:2012,Reines:2013} have, however,
shown that there is an offset at the low mass end of the $M_{\rm bh}$--$M_{\rm
  sph}$ diagram, such that the black hole mass is lower than predicted by the
near-linear $M_{\rm bh}$--$M_{\rm sph}$ relation established using galaxies
having predominantly higher mass black holes.  These offset galaxies have been
labelled by some to contain pseudobulges --- spheroidal components thought to
be produced by
the secular evolution of a disk and associated with bars 
\citep{Bardeen:1975,Hohl:1975,Hohl:1979,Combes:1981,Kormendy:1982,Kormendy:1993,KK:2004}. 
This would agree with one of 
the scenarios\footnote{Pseudobulges could be offset low in the $M_{\rm
    bh}$--$M_{\rm sph}$ diagram if secular evolution
disproportionately increases the central bulge mass relative to the growth
of the black hole.} presented by \citet{Hu:2008} and \citet{Graham:2008a}, and 
subsequently \citet{Kormendy:2011}, 
but if correct would present a contradiction with the
picture presented in the preceding paragraph.  Pseudobulges 
are, however, particularly difficult to reliably identify \citep{Wyse:1997}
because they can possess the same physical properties as low-mass, merger-built
bulges, including S\'ersic (1968) index, rotation, the presence of embedded discs,
and a systematic departure from the bright end of any scaling relation which
has used 'effective' radii or `effective' surface brightnesses
\citep[e.g.][and references
  therein]{Dom:1998,Aguerri:2001,Bekki:2010,Saha:2012,Querejeta:2014,Graham:2014b,Graham:2013}.

To address the above contradiction, and bypass the issue of pseudobulges, 
we start by noting that the near-linear scaling
relations between $M_{\rm bh}$ and host spheroid luminosity $L_{\rm sph}$,
and also host spheroid stellar mass $M_{\rm sph,*}$, 
\citep{Dressler:1989,Yee:1992,KR:1995,Magorrian:1998,Marconi:2003,Haring:2004}\footnote{The
linear $M_{\rm QSO}$--$M_{\rm galaxy}$ relationship proposed by
\citet{Yee:1992} pertains to the limit in massive spheroids, 
and it is effectively the relationship for which 
\citet{Magorrian:1998} and \citet{Laor:2001} later provided the zero-point.}
 have recently been shown by
\citet{Graham:2012a} to provide an incomplete description of the 
(black hole)--spheroid relationship.  In essence, the $M_{\rm bh} \propto \sigma^5$
\citep{FaM:2000,MaF:2001a,Graham:2011,McConnell:2011,GS:2013} and $L_{\rm sph}
\propto \sigma^2$ \citep{Davies:1983,Held:1992,Matkovic:2005,deRijcke:2005,
Balcells:2007b,Chilingarian:2008,Forbes:2008,Cody:2009,Tortora:2009,Kourkchi:2012}
scaling relations for ``S\'ersic spheroids'' necessitates a non-linear
$M_{\rm bh}$-$L_{\rm sph}$ and $M_{\rm bh}$-$M_{\rm sph,*}$ relation. 
S\'ersic spheroids are elliptical galaxies and the bulges
of disk galaxies 
which do not have partially depleted cores\footnote{See \citet{Graham:2005}
for a review of the S\'ersic model, and \citet{Graham:2003} for a description
of the core-S\'ersic model.}; they typically have $B$-band absolute
magnitudes $M_B \gtrsim -20.5\pm1$ mag, 
and S\'ersic indices $n \lesssim 3$--4. 
Graham (2012) pointed out that in these spheroids 
one expects to find that $M_{\rm bh} \propto L^{2.5}$ and that the
relationship between $M_{\rm bh}$ and $M_{\rm sph,*}$ should be better described by a
near-quadratic relation than a linear relation, as was further shown in 
\citet{GS:2013} and \citet{Scott:2013}.  As noted in these works, it is only
at high masses ($M_{\rm bh} \gtrsim 10^8 M_{\odot}$) that a near-linear
$M_{\rm bh}$-$M_{\rm sph,*}$ 
relation is evident, giving rise to this 'broken' scaling
relation.  Although, due to the scatter in the $M_{\rm bh}$--$M_{\rm sph}$ 
diagram, coupled with the location of the brighter S\'ersic galaxies at the high-mass end of the
near-quadratic $M_{\rm bh}$-$M_{\rm sph,*}$ relation, 
surveys which have not sufficiently probed below $M_{\rm bh} \approx 10^7 
M_{\odot}$ can readily miss the bend in the $M_{\rm bh}$--$M_{\rm sph}$ relation 
\citep[e.g.][]{Sani:2011,Beifiori:2012,Vika:2012,vandenBosch:2012,McConnell:2013,Sanghvi:2014,Feng:2014}. 
A steeper than linear, although not bent, relation was however detected 
early-on \citep{Laor:1998,Wandel:1999,Laor:2001}, and 
a number of recent theoretical works have now revealed a steepening 
relationship at lower masses 
\citep{Dubois:2012,Khandai:2012,Bonoli:2014,Bellovary:2014}, although
\citet[][their figure~22]{Khandai:2014} does not.
This would appear to be bringing things more in line with the prediction 
by \citet{Haehnelt:1998} that $M_{\rm bh} \propto M^{5/3}_{\rm halo}$. 

These scaling relationships are important for several reasons.  
Given the broken, or rather bent, $M_{\rm bh}$-$M_{\rm sph,*}$ 
relation, it implies that, within the S\'ersic galaxies, the supermassive
black holes grow more rapidly than the stellar spheroids \citep{Graham:2012a}.
That is, there is no tandem, lockstep, growth of black holes and bulges in
these galaxies: the 
$M_{\rm bh}$/$M_{\rm sph,*}$ mass ratio is not a constant value. 
Indeed, while this (previously thought to be constant) ratio was doubled in
\citet{Graham:2012a} for the massive galaxies, and then increased further 
to an {\it average} value of 0.49\% in the core-S\'ersic galaxies 
\citep[][see Laor 2001]{GS:2013}, it can be lower than $\sim 
10^{-3}$ in the lower-mass S\'ersic galaxies 
\citep[][see also Wandel 1999]{GS:2013,Scott:2013}\footnote{Applying dust
  corrections \citep[][see their figures~1 and 7]{GW:2008} to the bulge 
magnitudes of \citet{McLure:2001} will re-brighten the bulge luminosities of \citet{Wandel:1999}.}.
Low-mass 
spheroids therefore offer an even more promising domain, than previously
thought when assuming a constant $M_{\rm bh}$/$M_{\rm sph,*}$ mass ratio of
0.14--0.2\% \citep[][and references therein]{Ho:1999,MaF:2001b,Haring:2004}, 
to find new intermediate mass black holes, i.e.\ 
those with $10^2 < M_{\rm bh}/M_{\odot} < 10^5$. 

An additional, related, reason pertains to feedback from
supermassive black holes \citep[e.g.][]{Page:2012, Wurster:2013,
  Fanidakis:2013} which is commonly
thought to regulate star formation in the spheroid and provide a potential solution to the
over-abundance of massive galaxies predicted by dark-matter only simulations.
This process has been invoked to produce the turn-off in the galaxy luminosity
function at high luminosities 
(e.g.\ \citet{Benson:2003,Bower:2006,Croton:2006}). 
As noted, the near-quadratic $M_{\rm bh}$-$M_{\rm sph,*}$ relation for S\'ersic
spheroids flattens into a slope close to unity for the brighter 
core-S\'ersic galaxies \citep{Graham:2012a}. 
The presence of a 
partially-depleted core in these bigger spheroids is thought to indicate that
they and their black hole formed through simple, additive, dry major merger
events which created the near-linear (one-to-one) $M_{\rm bh}$-$M_{\rm sph,*}$ relation.
The process of ``mechanical'' or ``radio mode'' AGN feedback may 
therefore subsequently maintain, rather than establish, this linear 
relation\footnote{While this ``mechanical'' feedback may maintain the 
 (black hole)--spheroid relation, it might not necessarily prevent strong 
 accretion of a planar gas cloud and subsequent disk formation.}

Here we investigate if galaxies with AGN hosting low mass black holes 
that have been reported in the 
literature to be offset from the {\it near-linear} $M_{\rm bh}$-$M_{\rm
  sph,*}$ relation (defined by predominantly massive spheroids) might simply
be following the steeper relation of the S\'ersic galaxies.  If so, then they
may not be discrepant galaxies with unusually low $M_{\rm bh}$/$M_{\rm
  sph,*}$ mass ratios, but rather abide by the main relation defined by the
majority of galaxies today. 
This will have dramatic implications for cosmological hydrodynamical
simulations, such as Illustris \citep{Sijacki:2014}, which have tied
themselves to the near-linear $M_{\rm bh}$-$M_{\rm sph,*}$ relation. 
Our study has been performed by using data from many authors, thereby avoiding possible
biases in any one study, and deriving spheroid stellar masses when not done in
the original papers. 
In section~2 we introduce the galaxy/black hole samples used, and in Section~3
we present their location in the $M_{\rm bh}$-$M_{\rm sph,*}$ diagram. 
Section~4 provides a discussion of related topics such as formation theories, 
expectations for intermediate mass black holes, coexistence with nuclear star
clusters, and potential 
evolutionary pathways for possible under- and over-massive black holes. 

\section{Sample and Data}

\subsection{Reference sample} 

Our initial reference sample consists of 75 galaxies with directly measured
supermassive black hole masses and spheroid stellar mass
determinations\footnote{Some of the spheroid stellar masses used in this work
  are based on a {\it statistical} correction for the bulge-to-disk ratio
  rather than a direct measurement of the bulge luminosity and mass. For
  details of the correction and the additional scatter this is expected to
  introduce see \citet{GS:2013}.} \citep{Scott:2013}.  
The stellar masses were obtained by applying the ($B-K_s$)--( stellar
mass-to-light ratio) relation from \citet{Bell:2001} to the $K_s$-band
magnitudes of the 75 spheroids.  
The initial galaxy magnitudes came from the {\sc archangel} photometry pipeline
\citep{Schombert:2012} applied to Two Micron All-Sky Survey
\citep[2MASS][]{2MASS:2006} images.  
Inspection of the galaxy images (Savorgnan et
al., in prep.) has since resulted in 3 changes of morphological type: 
NGC~5845 (E $\rightarrow$ S0, with clear rotation seen by \citet{Emsellem:2011}); 
NGC~2974 (E $\rightarrow$ S0a, with faint spiral arms evident); 
and 
NGC~4388 (Sb $\rightarrow$ SBcd, due to a substantial edge-on bar that had inflated
past bulge estimates). 
This resulted in the following revised
$B-K_s$ colors and $K_s$-band magnitudes for their bulges:
(NGC~5845: 3.85, -21.84 mag); 
(NGC~2974: 3.66, -22.88 mag); and 
(NGC~4388: 3.72, -22.14 mag), and thus new spheroid masses following Eq.2 from
\citet{Scott:2013}. 
These are shown in Table~\ref{Tab1}. 
The presence, or otherwise, of a partially-depleted stellar core in the full
sample was primarily 
determined from high resolution Hubble Space Telescope imaging.  Here we
reclassify NGC~1332 and NGC~3998 as S\'ersic galaxies based on their light
profiles \citep{Rusli:2011,Walsh:2012}, whereas in \citet{GS:2013} and
\citet{Scott:2013} these galaxies had been tentatively classified as core-S\'ersic galaxies
based on their central velocity dispersion. 
In passing we note that NGC~1332 is a massive ($M=10^{11} M_{\odot}$), compact
($R_{\rm e} = 2$ kpc) early-type galaxy (Savorgnan et al.\ in prep.), with 
structural properties similar to NGC~1277\footnote{We do not include
NGC~1277 here because its black hole mass is currently being re-determined by
us and is expected to change dramatically from its published value.}.
Such galaxies, with properties similar to some at $z=2\pm0.5$, have been
claimed to be very rare in the Universe today \citep[e.g.][]{Trujillo:2014}.

We also now expand our reference sample through the addition of NGC~1316
(Fornax~A, $B-K_s=3.40$, M$_{K_s} = -24.73$ mag, following \citet{Scott:2013}). 
Although we had a measurement of its black hole mass and an 
estimate of its bulge magnitude in \citet{Graham:2012a}, we were uncertain 
as to whether it was a core-S\'ersic or a S\'ersic galaxy.  
Unbeknown to us at the time, 
\citet{Beletsky:2011} had however revealed the presence of a
kinematically cold, nuclear stellar disk with a radius less than 200 pc in
this peculiar, barred lenticular galaxy, expanding on the discovery by
\citet{D'Onofrio:2001}.  At odds with the initial classification in
\citet{Faber:1997}, NGC~1316 therefore does not posses a partially-depleted
stellar core.  

Finally, 
\citet{Licquia:2014} have provided a stellar mass estimate for the bulge of
the Milky Way, enabling us to also now include our own galaxy.  We have taken the
associated black hole mass from \citet{Chatz:2014}.  This gives us a final
sample of 77 galaxies with directly measured black hole masses.
 The changes since \citet{Scott:2013} that are mentioned above are captured in 
 Table~\ref{Tab1}.   It is also noted that four of the galaxies listed there have
 particularly high $M_{\rm bh}/M_{\rm sph}$ ratios.  Given the small degree of 
 change to the initial sample of 75 galaxies, we do not re-derive the $M_{\rm bh}$--$M_{\rm sph}$ scaling
 relations given in \citet{Scott:2013} for the core-S\'ersic and S\'ersic
 galaxies.  It is however noted that the denominator in equation~4 from 
 \citet{Scott:2013}, pertaining to the S\'ersic galaxies, contains a typographical error
 and should read $2\times10^{10}$ rather than $3\times10^{10}$.

\begin{table}[t]
\caption{Updates to the ($M_{\rm bh}, M_{\rm sph,*}$) table in \citet{Scott:2013}.}
\label{Tab1}
\begin{center}
\begin{tabular}{l c c c c}
\hline
Name      & Type &  $M_{\rm sph,*}$     & $M_{\rm bh}$               & Core \\
          &      & ($10^{9}$ M$_\odot$) & ($10^8$ M$_\odot$)         &      \\
\hline
\multicolumn{5}{c}{Revised core-S\'ersic/S\'ersic classification} \\
NGC 1332  & S0   & 47$^{+69}_{-28}$     & 15$^{+2}_{-2}$             & n \\
NGC 3998  & S0   & 14$^{+21}_{-9}$      & 8.1$^{+2.0}_{-1.9}$        & n \\
NGC 1316  & S0   & 9$^{+12}_{-5}$       & 1.5$^{+0.75}_{-0.8}$       & n \\
\hline
\multicolumn{5}{c}{New or revised spheroid mass} \\
NGC 2974  & S0a  &  19$^{+24}_{-11}$    &  1.7$^{+0.2}_{-0.2}$       & n \\
NGC 4388  & Scd  &  10$^{+13}_{-6}$     &  0.075$^{+0.002}_{-0.002}$ & n? \\
NGC 5845   & S0   &  8$^{+10}_{-4}$      &  2.6$^{+0.4}_{-1.5}$       & n \\
Milky Way & SBbc & 9.1$^{+0.7}_{-0.7}$  &  0.0426$^{+0.0014}_{-0.0014}$ & n \\
\hline
\hline
\end{tabular}
\end{center}
\end{table}

\subsection{Under-massive black hole candidates}

Our sample of galaxies in Tables~\ref{Tab2} and \ref{Tab3} 
with allegedly under-massive black holes, at least relative to
the single, near-linear $M_{\rm bh}$-$M_{\rm sph,*}$ relation defined by
predominantly massive galaxies, has come from the following papers. 

\citet{Jiang:2013} report that UM~625 falls below the $M_{\rm bh}$-$L_{\rm
  sph,*}$ relation.  They report a virial black hole mass of $1.6\times10^6
M_{\odot}$, a $V$-band bulge magnitude of $-$19.06 mag (accounting for 60\% of
this S0 galaxy's light), and a $V$-band stellar mass-to-light ratio of 1.6
$M_{\odot}/L_{\odot}$ which yields a bulge stellar mass of $5.4\times10^9 M_{\odot}$.

\citet{Yuan:2014} report on virial masses for 
two elliptical galaxies\footnote{\citet{Yuan:2014} include two additional
  galaxies with million solar mass black holes, but no bulge/disc
  decomposition is available for them.}:  
SDSS J004042.10−110957.6 ($M_{\rm bh}=1.22\times10^6 M_{\odot},
M_{sph,*}=11.8\times10^9 M_{\odot}$) and 
J074345.47+480813.5 ($M_{\rm bh}=0.51$--$0.66\times10^6 M_{\odot}, 
M_{sph,*}=20.5\times10^9 M_{\odot}$), assuming here that 
$M/L_B = 6\pm2.5$ \citep{Worthey:1994} 
and using $M_{\odot,B}=5.47$ mag \citep{Cox:2000}.

\citet[][their tables~1 and 6]{Reines:2013} present stellar masses along with 
black hole mass estimates derived using the broad H$\alpha$ line for ten 
dwarf galaxies, as done by \citet{Yuan:2014}.  
This emission-line method has been calibrated against reverberation mapping
techniques that require knowledge of the `virial factor' \citep{PaW:2000,
  Onken:2004}.  Using a common `forward regression' analysis,
\citet{Graham:2011} derived a virial factor of $3.8^{+0.7}_{-0.6}$ (cf.\
$5.1^{+1.5}_{-1.1}$ from \citet{Park:2012}) using a large sample of galaxies.
However \citet{Graham:2011} pointed out a 
sample selection bias (as opposed to a real / natural boundary) 
that misses low mass black holes, and which is reduced 
when using an `inverse regression' analysis, resulting in a virial factor
of $2.8^{+0.7}_{-0.5}$ (cf.\ $3.4^{+1.2}_{-0.9}$ from \citet{Park:2012}).  
For several reasons discussed in \citet{Graham:2011}, this value is an upper
limit; but it does (coincidentally?) 
agree with the isotropic spherical virial coefficient of 
3 from Netzer (1990). 
As noted in footnote~1 of \citet{Graham:2011}, 
due to differing notation in the literature, this value is sometimes reduced 
by a factor 4, and given as 0.75.   
While \citet{Jiang:2013} and \citet{Yuan:2014} used a `reduced' virial factor of 0.75, 
\citet{Reines:2013} used a value of 1.0.  For consistency we
have therefore reduced the black hole mass estimates of 
\citet{Reines:2013} by 1.0/0.75.  

For the dwarf Seyfert 1 galaxy POX~52 ($M_{\rm sph,*}=1.2\times10^9$), 
\citet{Thornton:2008} derived 
a black hole mass estimate using the radius-luminosity relation of 
\citet{Kaspi:2000} (giving a broad line region radius from the AGN luminosity) 
together with the velocity obtained from the 
broad H$\beta$ line width (enabling an $f^{\prime}.V^2R/G$ virial mass estimate). 
Adjusting the `reduced' virial factor $f^{\prime}$ that they used from 1.4
\citep{Onken:2004} to 0.75 \citep{Graham:2011}, 
lowers their virial mass estimate from (3.1--4.2)$\times10^5 M_{\odot}$ to 
(1.7--2.3)$\times10^5 M_{\odot}$. 

\citet{Mathur:2012} have presented ten black hole masses obtained 
via the virial relation of \citet{Kaspi:2000}, who used a `reduced' virial factor of
0.75, and we include this data set.  The absolute $r$-band bulge magnitudes presented
in \citet{Mathur:2012} were converted into a stellar mass using $M/L_r =
3.5\pm 1.5$ \citep{Worthey:1994} and $M_{\odot,r}=4.50$ mag (Vega).   
While the redshifts ($z$) of the galaxies from the four previously mentioned
studies in this section are all less than 0.05, and $z \le 0.04$ for all but
two of them, i.e.\ most are within 170 Mpc, eight of the ten galaxies from
\citet{Mathur:2012} have $z > 0.06$ and they reach out to $\sim$0.17.  
The $K$-correction for this sample is expected to be $\lesssim$0.1 
mag in the $r$-band \citep{Chilingarian:2010} and is therefore not bothered
with, especially as we do not have useful colour information for much of this
sample.  While
the higher redshift of 0.17 corresponds to 2.5$\log(1+z)^2 = 0.34$ mag of
cosmological redshift dimming of the observed magnitude, 
it somewhat cancels with the expected
evolutionary correction due to galaxies being brighter when they were younger, 
which is estimated to be 
$-1.2z$ mag for elliptical and Sc galaxies, and $-1.75z$ mag for Sa galaxies
\citep{Poggianti:1997}.  Due to this expected cancellation, coupled with our
uncertainty as to the morphological type, we have not applied these
corrections, but instead note that there could be a tenth or a couple of
tenths of a magnitude error because of this.  

\citet{Busch:2014} report on a sample of 11 low-luminosity type-1 quasars whose
black hole masses reside below the linear $M_{\rm bh}$--$M_{sph,*}$ relation.
Their figure~14 reveals that the location of their data in the $M_{\rm
  bh}$--$M_{sph,*}$ diagram overlaps with the distribution from the
large, predominantly inactive galaxy sample used to define the bent $M_{\rm
  bh}$--$M_{sph,*}$ relation in \citet{Scott:2013}.  
Here we reduce their black hole masses by 3/3.85 as they used a virial factor
of 3.85 from \citet{Collin:2006}.  
We have derived the spheroid masses from the
absolute $K$-band magnitudes (which are minimally affected by dust) 
listed in their table~7, by using an average 
$M/L_K = 0.8$ and $M_{\odot ,K} = 3.28$ mag (Vega). \citet{Busch:2014} reported a
range of $M/L_K$ values from 0.73 to 0.85, slightly greater than the typical
value of 0.6 reported by \citet{MaS:2014} for disk galaxies. 
Given that the galaxy sample from \citet{Busch:2014} has $z \lesssim 0.06$, any 
cosmological corrections would be smaller than the 0.2 mag 
uncertainty on the magnitudes reported by \citet{Busch:2014}. 
However the dominant uncertainty may well be in the conversion from 
stellar light to stellar mass. 

As noted by \citet{Busch:2014}, 
if a significant fraction of young stars is present, then our adopted 
$M/L$ ratio is too high and we have over-estimated the spheroid mass. 
The same situation may occur with the sample from \citet{Yuan:2014}
and \citet{Mathur:2012} (which is also partly why we are not 
particularly concerned with a possible $\sim$0.1--0.2 mag error in their
bulge magnitudes). 
Given the co-existence of AGN and star formation, due to the available gas
supply, it seems plausible that this could be the case 
\citep[][his sections 2 and 3]{Alexander:2012}. 
Although in determining this, one 
obviously needs to distinguish between star formation contributing toward
bulge versus disk growth, and 
be aware that a large fraction of BH accretion, and star formation
\citep{Straatman:2014}, is obscured by dust \citep{Webster:1995,Del:2013,Assef:2014}. 
Obtaining better stellar mass-to-light ratios are, thankfully, a topic already 
under investigation by Busch (priv.\ comm.). 
The masses for our sample from \citet{Busch:2014}, plus all the above masses, 
are collated in Table~\ref{Tab2} to give a total of 35 AGN.

\begin{table*}
\caption{``Under-massive'' black hole sample}
\label{Tab2}
\begin{center}
\begin{tabular}{@{}lcccc@{}}
\hline
Galaxy & $M_{\rm bh}$     & Mag$_{\rm sph}$  &  M/L &  $M_{\rm sph,*}$ \\
       & $10^{5} M_\odot$ &     [mag]          &      &  $10^9 M_\odot$  \\
\hline
\multicolumn{5}{c}{\citet{Thornton:2008}} \\
Pox~52 & 2.0$\pm$0.3     &  ...                & ...  & 1.2   \\
\hline
\multicolumn{5}{c}{\citet{Jiang:2013}} \\ 
UM 625 & 16              & $-$19.06 $V$-mag    &  1.6 & 5.4  \\
\hline
\multicolumn{5}{c}{\citet{Yuan:2014}} \\
SDSS J004042.10-110957.6  & 12.2       & $-$17.4 $B$-mag  & 6$\pm$2.5  &  8.4$^{+3.6}_{-3.5}$  \\ 
SDSS J074345.47+480813.5  & 5.1--6.6   & $-$18.0 $B$-mag  & 6$\pm$2.5  &  14.7$^{+6.1}_{-6.2}$  \\ 
\hline
\multicolumn{5}{c}{\citet{Reines:2013}, BPT AGNs} \\
SDSS J024656.39-003304.8     & 5.0  &  ... & ... &       2.57  \\ 
SDSS J090613.75+561015.5     & 2.5  &  ... & ... &       2.29  \\ 
SDSS J095418.15+471725.1     & 0.8  &  ... & ... &       1.32  \\ 
SDSS J122342.82+581446.4     & 12.6 &  ... & ... &       2.95  \\ 
SDSS J122548.86+333248.7     & 1.0  &  ... & ... &       1.26  \\ 
SDSS J144012.70+024743.5     & 1.6  &  ... & ... &       2.88  \\
\multicolumn{5}{c}{\citet{Reines:2013}, BPT Composites} \\
SDSS J085125.81+393541.7     & 2.5  &  ... & ... &       2.57  \\
SDSS J152637.36+065941.6     & 5.0  &  ... & ... &       2.14  \\ 
SDSS J153425.58+040806.6     & 1.3  &  ... & ... &       1.32  \\ 
SDSS J160531.84+174826.1     & 1.6  &  ... & ... &       1.74  \\ 
\hline
\multicolumn{5}{c}{\citet{Mathur:2012}} \\
TON S180                 &  71  & $-$20.1 $r$-mag  & 3.5$\pm$1.5  &   24.2$^{+10.4}_{-10.4}$ \\
RX J1117.1+6522          & 210  & $-$19.7 $r$-mag  & 3.5$\pm$1.5  &   16.8$^{+7.1}_{-7.2}$ \\
RX J1209.8+3217          &  54  & $-$19.8 $r$-mag  & 3.5$\pm$1.5  &   18.4$^{+7.8}_{-7.9}$ \\
IRAS F12397+3333         &  45  & $-$20.2 $r$-mag  & 3.5$\pm$1.5  &   26.6$^{+11.3}_{-11.4}$ \\
MRK 478                  & 269  & $-$21.2 $r$-mag  & 3.5$\pm$1.5  &   66.7$^{+28.6}_{-28.6}$ \\
RX J1702.5+3247          & 217  & $-$19.8 $r$-mag  & 3.5$\pm$1.5  &   18.4$^{+7.8}_{-7.9}$ \\
RX J2216.8-4451          & 167  & $-$21.1 $r$-mag  & 3.5$\pm$1.5  &   60.8$^{+26.1}_{-26.0}$ \\
RX J2217.9-5941          & 124  & $-$19.6 $r$-mag  & 3.5$\pm$1.5  &   15.3$^{+6.5}_{-6.6}$ \\
MS 2254.9-3712           &  39  & $-$19.1 $r$-mag  & 3.5$\pm$1.5  &   9.64$^{+4.2}_{-4.1}$ \\
MS 23409-1511            & 100  & $-$20.7 $r$-mag  & 3.5$\pm$1.5  &   42.1$^{+18.0}_{-18.1}$ \\
\hline
\multicolumn{5}{c}{\citet{Busch:2014}} \\
HE0045-2145              &  6.0 & $-$22.42 $K_s$-mag & 0.8        &  15.2  \\
HE0103-5842              & 39.8 & $-$23.78 $K_s$-mag & 0.8        &  53.3  \\
HE0224-2834              & 331  & $-$24.48 $K_s$-mag & 0.8        &  102   \\
HE0253-1641              & 39.8 & $-$22.13 $K_s$-mag & 0.8        &  11.7  \\
HE1310-1051              & 166  & $-$23.25 $K_s$-mag & 0.8        &  32.7  \\
HE1338-1423              & 155  & $-$23.97 $K_s$-mag & 0.8        &  63.5  \\
HE1348-1758              & 15.5 & $-$21.99 $K_s$-mag & 0.8        &  10.3  \\
HE1417-0909              & 135  & $-$23.05 $K_s$-mag & 0.8        &  27.2  \\
HE2128-0221              & 195  & $-$23.36 $K_s$-mag & 0.8        &  36.2  \\
HE2129-3356              & 490  & $-$23.23 $K_s$-mag & 0.8        &  32.1  \\
HE2204-3249              & 1000 & $-$25.00 $K_s$-mag & 0.8        &  164   \\
\hline
\end{tabular}
\end{center}
\end{table*}

\begin{figure}
  \includegraphics[scale=0.46,angle=-90]{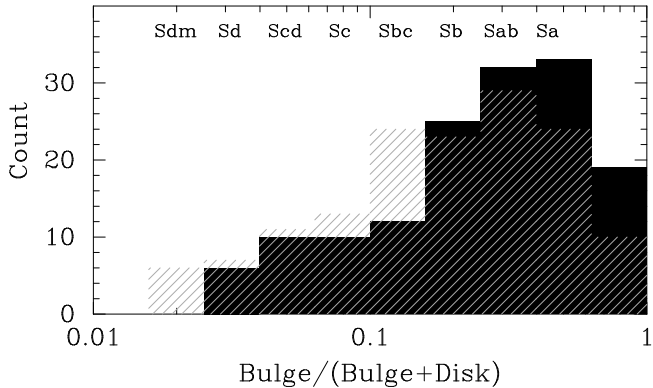}
  \caption{Bulge-to-(bulge+disk) flux ratios for 147 galaxies from 
\citet{Jiang:2011a}.  The grey hashed histogram shows the ratios after correcting 
for Galactic dust extinction, while the solid histogram shows the ratios after
additionally applying the internal galaxy dust corrections from \citet{Driver:2008}. 
This can be compared with figure~5 from \citet{Jiang:2011b}.}
  \label{Fig1}
\end{figure}

Finally, we use the large data set from \citet{Jiang:2011a,Jiang:2011b}, 
providing an additional 147 virial black 
hole mass estimates (obtained using a `reduced' virial factor $f^{\prime}=0.75$) 
and $I$-band bulge and disk magnitudes. 
Given that this is our largest data set, we 
dedicate some space to describing our conversion of their 
published magnitudes into stellar masses. 
We have applied the following five corrections to the apparent $F814W$ ($I$-band) 
bulge magnitudes. 
(i) They are corrected for foreground Galactic extinction
using the recalibrated dust extinction maps of \citet{Schlafly:2011} as given
in NED\footnote{NASA/IPAC Extragalactic Database: http://ned.ipac.caltech.edu/}, 
and 
(ii) further
brightened by 2.5$\log(1+z)^2$ due to cosmological redshift ($z$) dimming.
(iii) 
We then correct the Sloan Digital Sky Survey (SDSS) Data Release 6 \citep[DR6,][]{Adelman:2008} 
$g-i$ color of each galaxy (available through NED) for Galactic dust, 
enabling us to use the (foreground extinction)-corrected, ($g-i$)-based 
$K$-corrections from \citet{Chilingarian:2010}. 
The corrections obtained apply to the SDSS $i$-band but were assumed to be 
suitably applicable to the HST $I$-band. 
These $K$-corrections are small, with all but 3 (29) galaxies requiring 
an adjustment smaller than 0.2 (0.1) mag.  
(iv) The bulge and disk magnitudes are then separately corrected for internal dust
extinction using the generic $i$-band formula given in \citet{Driver:2008} and assuming the
reported axis ratios (courtesy of NED) reflect the inclination of each
galaxy's disc. 
(v) The final correction to the magnitude is to evolve the bulges to $z=0$.
To do this, the bulge-to-disk flux ratio is calculated and used to estimate the 
morphological type based on the dust-corrected bulge-to-disk flux ratios given in table~8
of \citet{GW:2008}.  Figure~\ref{Fig1} shows the results. 
Although the sample is dominated by early-type (Sb and earlier) 
disk galaxies, 26 late-type (Sc and later) galaxies are present. 
We have then roughly 
applied the (morphological-type)-based $I$-band evolutionary corrections
from \citet{Poggianti:1997} by using a redshift correction of $-1.6z$ for the
Sa ($0.4 < (B/T) < 0.6$) galaxies and $-z$ for the remainder.  
Given that all but 6 (1) galaxies have a redshift less than 0.2 (0.35), this
is also a small overall correction, which is fortunate given the
uncertainties associated with this particular correction for disk galaxies. 
The corrected apparent bulge magnitudes are then converted into absolute
magnitudes assuming $H_0=70$ km s$^{-1}$, $\Omega_m =0.3$ and 
$\Omega_{\Lambda}=0.7$. 

Overlooking the small difference between $i$ and $I$ magnitudes, 
we use the relation between $M_*/L_i$ and $g-i$ color in \citet{Taylor:2011},
such that $\log(M_*/L_i)=-0.68+0.70(g-i)$, to convert the bulge
magnitudes into stellar masses. 
This relation was calibrated for galaxies with masses down to a few times
$10^8 M_{\odot}$, which essentially covers the full range of the \citet{Jiang:2011a} sample.
Our $g-i$ color was corrected for Galactic extinction, as noted above, and
also $K$-corrected.  It was not corrected for dust nor evolution.  
This resulted in $M_*/L$ ratios typically ranging from 0.55 to 1.15. 
We excluded 8 galaxies because two had no $g-i$ color (1033+6353 and 1127+4625),
the $z=0.614$ galaxy 1253+4627 had an unknown $K$-correction, and five had
far too red $g-i$ colors around 1.4$\pm$0.04 (0347+005, 0927+0843, 1027+4850,
1153+5256, 1621+3436).  The resulting masses for the remaining 139 bulges 
are given in Table~\ref{Tab3}.

\begin{table*}
\caption{``Under-massive'' black hole sample from \citet{Jiang:2011a}}
\label{Tab3}
\begin{center}
\begin{tabular}{@{}lcccccccccc@{}}
\hline
Galaxy                  & $z$    & $\log(M_{\rm bh}/M_{\odot})$ & $m_{I,orig}$ & $A_i$ & $A_g$ & $g-i$ & $M/L$ & $b/a$ & $m_{I,corr}$ & $\log(M_{\rm sph,*}/M_{\odot})$ \\
 (1)                    & (2)    &   (3)                    &  (4)             &  (5)  &  (6)  &  (7)  &  (8)  &  (9)  &  (10)      &   (11)  \\ 
\hline                                                                                                                                  
SDSSJ002228.36-005830.6 & 0.1060 &   5.7                    & 19.40 $\pm$ 0.37 & 0.045 & 0.088 & 0.585 & 0.44  & 0.83 & 18.57 & 9.25 \\ 
SDSSJ002452.53-103819.6 & 0.1030 &   6.2                    & 18.93 $\pm$ 0.10 & 0.051 & 0.100 & 1.185 & 1.07  & 0.91 & 18.30 & 9.73 \\ 
SDSSJ011749.81-100114.5 & 0.1410 &   5.8                    & 18.40 $\pm$ 0.03 & 0.069 & 0.134 & 0.923 & 0.70  & 0.89 & 17.67 & 10.09 \\ 
SDSSJ012055.92-084945.4 & 0.1250 &   6.3                    & 19.28 $\pm$ 0.02 & 0.070 & 0.136 & 1.132 & 0.94  & 0.67 & 18.48 & 9.77 \\ 
SDSSJ015804.75-005221.8 & 0.0804 &   5.9                    & 17.42 $\pm$ 0.03 & 0.044 & 0.085 & 0.889 & 0.75  & 1.00 & 16.84 & 9.92 \\ 
\hline
\end{tabular}

\noindent
Notes: The first 3 columns have been taken from table~1 in \citet{Jiang:2011a}, and show
the galaxy name, its redshift and virial black hole mass estimate. 
The fourth column shows the $i$-band bulge magnitude as reported in table~2 of \citet{Jiang:2011a}. 
The fifth and sixth columns show the $i$- and $g$-band Galactic extinction
taken from NED, as described in the text. 
The $g-i$ color shown in column 7 has come from {\it SDSS DR6}, 
but has been corrected here for Galactic extinction in the $g$ and $i$ bands. 
This was used to obtain 
the mass-to-light ($M/L$) ratios shown in column 8 and obtained from
equation~7 in \cite{Taylor:2011} assuming 
$M_{i,\odot} = 4.15$ mag (Vega).  Column 9 shows the minor-to-major axis ratio. 
Column 10 shows our final, corrected apparent bulge magnitudes (see the text for full details). 
Finally, the stellar mass of the bulge is given in column 11. 
Only a portion of this table is shown here to demonstrate its content.
\end{center}
\end{table*}

\section{Results}
\label{sec:results}

\begin{figure}
  \includegraphics[scale=0.43,angle=-90]{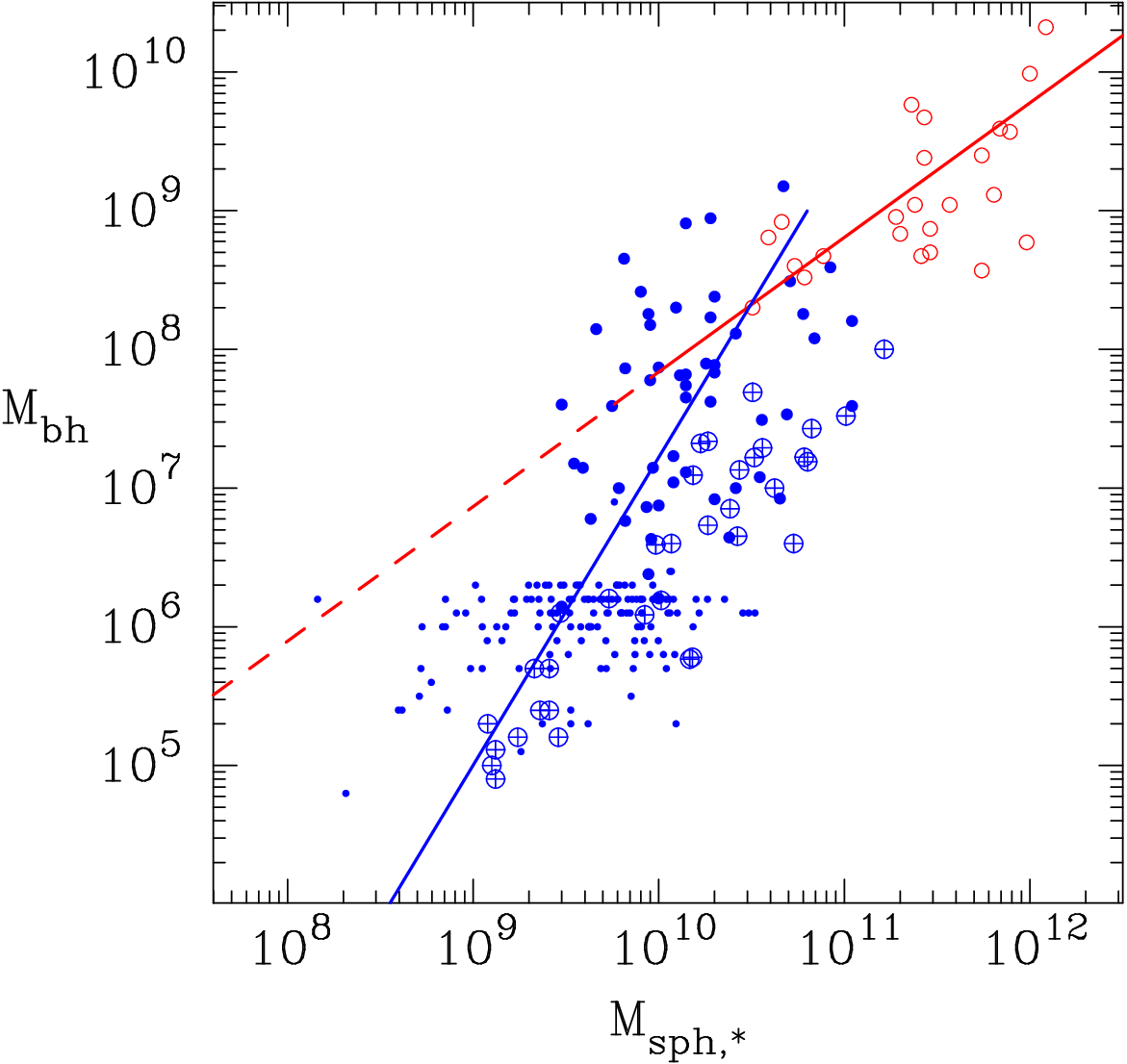}
  \caption{$M_{\rm bh}$ versus $M_{\rm sph,*}$, in units of solar masses, 
    for the sample of 77 galaxies 
    with direct $M_{\rm bh}$ measurements from \citet{Scott:2013}, plus
    the (147-8=139) low-mass AGN sample from \citet{Jiang:2011a} (small dots) 
    and the 35 additional galaxies (cross-hairs) listed in Table~\ref{Tab2}. 
    The S\'ersic galaxies from \citet{Scott:2013} are shown by the 
    filled blue circles, with the core-S\'ersic galaxies denoted by open red
    circles (slightly updated here according to Table 1).  
    The near-linear and near-quadratic scaling relations from \citet{Scott:2013} 
    are shown as the red (solid and dashed) and blue (solid) line for the 
    core-S\'ersic and S\'ersic galaxies, respectively. 
    The deviant spheroid with the lowest mass is SDSS 0840+4123.}
  \label{Fig2}
\end{figure}

In Figure~\ref{Fig2} we expand upon figure~3 from \citet{Scott:2013}, showing
the $M_{\rm bh}$--$M_{\rm sph,*}$ distribution for 77 galaxies with directly
measured black hole masses, divided into S\'ersic galaxies (blue filled
circles) and core-S\'ersic galaxies (red open circles).  The best-fitting
log-linear relations from \citet{Scott:2013} for these two sub-samples are
shown, with the extrapolation of the near-linear core-S\'ersic relation 
(slope = 0.97$\pm$0.14) to low masses
shown with the dashed line to highlight the difference with the steeper
S\'ersic relation (slope = 2.22$\pm$0.58). 
By presenting spheroid masses for an additional $35+139$ AGN with indirect
black hole mass determinations, and combining data from several papers, 
it becomes apparent
that these additional galaxies are not randomly distributed below the linear
$M_{\rm bh}$--$M_{\rm sph,*}$ relation for the core-S\'ersic 
galaxies but instead appear to somewhat overlap with 
the S\'ersic sequence of galaxies that have S\'ersic indices ranging from less
than 1 in small spheroids up to 3--4 in the larger spheroids. 

Claims for increased scatter at the low mass end of the 
$M_{\rm bh}$--$M_{\rm sph,*}$ relation, or even a breakdown in
this relation, may be misleading if one overlooks that the slope of these
relations steepen here.  If sampling too small a range in spheroid mass or black hole
mass, one may of course also fail to recover the $M_{\rm bh}$--$M_{\rm sph,*}$ relation. 
Due to the small range in black hole mass in the sample from 
\citet{Jiang:2011a}, of their 139 galaxies that we could use, a bisector
regression yields a slope consistent with zero: 0.12$\pm$0.25. 
We can however use this large homogeneous, and our most carefully treated data
sample, to investigate the scatter in the $\log M_{\rm sph,*}$ direction. 

The median horizontal offset of the \citet{Jiang:2011a} data from the
near-quadratic $M_{\rm bh}$--$M_{\rm sph,*}$ relation given by
\citet{Scott:2013} is just 0.19 dex, and an offset of zero is obtained  
by adjusting the slope of that relation within the 1$\sigma$ unceratinty
quoted by \citet{Scott:2013}. 
By contrast, the median horizontal offset about the near-linear 
$M_{\rm bh}$--$M_{\rm sph,*}$ relation from \citet{McConnell:2013} is 1.63
dex, i.e.\ a factor in excess of 40.

If the scatter in the $M_{\rm bh}$--$M_{\rm sph,*}$ diagram remains constant
in the horizontal ($\log M_{\rm sph}$) direction, then the scatter in the
vertical ($\log M_{\rm bh}$) direction will naturally increase where the
relation steepens.  Looking at the \citet{Jiang:2011a} data in
Figure~\ref{Fig2}: relative to the S\'ersic relation (after accounting for the
mean 0.19 dex displacement of the \citet{Jiang:2011a} data to 
higher spheroid masses), 
68\% of their data (i.e.\ +/-34\%) is contained
within 0.83 dex in the horizontal direction.  That is, their galaxy sample has
a 1$\sigma$ scatter of $\sim$0.42 dex in the horizontal direction about the 
near-quadratic S\'ersic $M_{\rm bh}$--$M_{\rm sph,*}$ relation.  This level of
scatter is comparable with the level of scatter commonly reported in the
vertical direction around the near-linear segment of the $M_{\rm bh}$--$L_{\rm
  sph,*}$ and $M_{\rm bh}$--$M_{\rm sph,*}$ 
relation defined by the bright spheroids.  The scatter in spheroid
mass, at a given black hole mass, therefore appears to be similar at the low- and
high-mass end of the $M_{\rm bh}$--$M_{\rm sph,*}$ diagram.
At the low mass end, for a slope of 2, the 1$\sigma$ scatter should thus be
0.83 dex in the vertical direction.  
If, at these low masses, the bulk of the data reside within $\pm2\sigma$ of the 
near-quadratic relation, then the 
observed range in black hole mass at a given spheroid mass should be 3.32
dex. It is therefore not surprising that studies with a limited range in spheroid
mass (as opposed to black hole mass) may also miss detecting the relation. 

In spite of the many sources of scatter that our remaining, heterogeneous AGN
sample (Table~\ref{Tab2}) 
may contain, these AGN appear to follow a sequence; a sequence whose slope is steeper
than 0.97 and less than 2.22.  However, given the previously mentioned possibility that we
may have over-estimated the mass-to-light ratios, and thus the bulge masses, of
some of our AGN hosts, we feel that it would be premature to place too much
confidence in a line fit to this data.  Although, our distribution of spheroid stellar
masses for the 139 AGN from \citet{Jiang:2011a} does however appear broadly
consistent with the distribution of dynamical ($5\sigma^2\,R_{\rm e}$) masses
shown for some of these galaxies by \citet[][their Figure~1b]{Jiang:2011a}.
This lends support to both our, and their, mass estimates.  Unfortunately we
do not have dynamical mass estimates to compare with for the other AGN samples
which are marked with a different symbol in Figure~\ref{Fig2}.  Having noted our
concerns, applying a bisector regression to these 35 AGN, they are found to have a slope of
1.56$\pm$0.10 in Figure~\ref{Fig2} (obtained using a simple factor of 
2 error for all spheroid and black hole masses).  
This slope is shallower than 2.22, but it does have overlapping 1$\sigma$ error
bars with the measurement 2.22$\pm$0.58. 
In summary, the AGN are not randomly 
offset; they follow a steeper relation than the near-linear (i.e.\ slope
close to 1) relation defined by the massive systems.

\section{Discussion}

The magnitude or mass marking the divide between the S\'ersic and the
core-S\'ersic galaxies --- which experienced different evolutionary paths ---
is fairly broad.  In Figure~\ref{Fig2} this transition occurs over the mass
range $M_{\rm sph,*} = 3\times10^{10} M_{\odot}$ to $10^{11} M_{\odot}$.  This
transition has been associated with the change in slope of the $L_{\rm
  sph}$--$\sigma$ relation \citep[e.g.][and references
  therein]{Matkovic:2005,Graham:2013}.  Indeed, the transition across the bend
in the $M_{\rm sph}$--$\sigma$ diagram from \citet[][their
  figure~1]{Cappellari:2013} matches this same mass range \citep{Korea:2014}.
It is also interesting to recall that \citet{Laor:2000} remarked that only
AGNs with black hole masses $\gtrsim 2\times 10^8 M_{\odot}$ --- which
corresponds to the onset of the transition in Figure~\ref{Fig2} --- generate
large scale jets, presumably capable of halting star formation and maintaining
the near-linear $M_{\rm bh}$--$M_{\rm sph,*}$ relation seen at high masses.

This division can be related to galaxy structure, in addition to the galaxy
dynamics.  There is a log-linear relation between the luminosity and the
central surface brightness ($\mu_0$) of elliptical galaxies from $M_B \approx
-14$ mag to $\approx -20.5$ mag \citep{JaB:1997}.  
There is also a log-linear relation between
the luminosity and the S\'ersic index of these galaxies \citep{GaG:2003}.  The lack of a break
in the above two log-linear relations at $M_B \approx -18$ mag (S\'ersic $n
\approx 2$) unites the faint and intermediate galaxies\footnote{As a result of
  these two linear relations, diagrams using ``effective'' radii and
  ``effective'' surface brightnesses are predicted to be (and found to be)
  curved.  They lead to the false impression of a division at $M_B \approx
  -18$ mag \citep[][and references therein]{Graham:2013}.}.  However galaxies
brighter than $M_B \approx -20.5 (\pm0.75)$ mag, with partially depleted
cores, deviate from the $L$--$\mu_0$ relation established by the fainter
galaxies (Graham \& Guzm\'an 2003).  For an old stellar population with $M/L_B
= 8$, $M_B = -20.5 (\pm0.75)$ mag corresponds to a mass of
$2^{+2}_{-1}\times10^{11} M_{\odot}$, which is where the core-S\'ersic
galaxies start to dominate the $M_{\rm bh}$--$M_{\rm sph,*}$ diagram.  While
these core-S\'ersic galaxies can have large scale discs
\citep[e.g.][]{Dullo:2013}, at $M_{\rm sph} \gtrsim 4\times10^{11} M_{\odot}$
they tend to be slow rotators \citep{SAURON:2007}.

The lower mass host galaxies of the AGN with low mass black holes
are of course not expected to have partially depleted cores like the giant
core-S\'ersic galaxies.  For this reason, it is not surprising that they
follow the same relation as the S\'ersic galaxies in the $M_{\rm bh}$--$M_{\rm
  sph,*}$ diagram.

Related to the S\'ersic versus core-S\'ersic separation, 
\citet{GS:2013} flagged two main regimes of black hole growth: 
gas-dominated processes occurring in S\'ersic galaxies and gas-poor (dry)
major merging forming the core-S\'ersic galaxy sequence in Figure~\ref{Fig2}. 
In this scenario, the low-mass spheroids grow their black 
holes rapidly, relative to the spheroid, through the accretion of gas and
stellar material (possibly also merging with other supermassive black holes).
Supporting this scenario, \citet{Gabor:2013} have recently demonstrated that
black holes grow more rapidly, relative to their host spheroid, in lower-mass
and gas-rich galaxies. Indeed, a tide of papers
\citep{Diamond:2012,Seymour:2012,Agarwal:2013,LaMassa:2013,Lehmer:2013,Drouart:2014} now
reveal changes in the (black hole)-to-galaxy mass ratio which may support such
growth (see also \citet{Trakhtenbrot:2012} and \citet{Alonso:2013}).
For example, \citet{LaMassa:2013} report that the star formation growth is
related to the black hole growth raised to the power of 0.36.  Flipping this
implies $M_{\rm bh}\propto M_{\rm sph,*}^{2.78}$.  

In the low-mass regime, the growth of the black hole and spheroid is linked
because they both grow from the same source, the galaxy's cold gas
reservoir. This apparently establishes the quadratic or
``super-quadratic''\footnote{The term ``super-quadratic'' is used to denote an
  exponent in a scaling relation that is greater than 2 but less than 3.}
relation between the black hole mass and that of the host spheroid during the
quasar's ``radiative mode'' \citep{GS:2013}.  This continues until a critical
point is reached, around a spheroid stellar mass of 
0.3--1.0$\times10^{11}\ M_\odot$.  At this mass, radio-mode feedback, also known as
``mechanical mode'' feedback, from the black hole may become effective enough
to expel the majority of the galaxy's cold gas reservoir and prevent this gas
cooling again \citep{Silk:1998, Haehnelt:1998}, as implemented in many
semi-analytical/numerical codes 
\citep{Kawata:2005,Springel:2005,Bower:2006,Merloni:2008,Booth:2009}.  
However, this latter feedback is not 
actually responsible for establishing the scaling relation between $M_{\rm
  bh}$ and the host spheroid.  With little gas reservoir to accrete from, the
supermassive black hole now grows predominantly through dry merging with other
massive black holes, leading to the core-S\'ersic relation with linear growth
of the black hole and host spheroid, maintained by the black hole's
``mechanical/radio mode'' feedback \citep{Karouzos:2014}, and / or perhaps
also from super stellar winds \citep{Conroy:2014}. 

The clever, many-merger scenario proposed by \citet{Peng:2007}, see
  also \citet{Jahnke:2011, Hirschmann:2010}, to produce a linear one-to-one scaling
via the central limit theorem can be ruled out.  Using a sample of galaxies
with a range of initial $M_{\rm bh}/M_{\rm gal,*}$ mass ratios,
\citet{Peng:2007} noted that after many mergers it will create an $M_{\rm
  bh}$--$M_{sph,*}$ relation with a slope of 1.  This idea was attractive when
it was thought that a single linear $M_{\rm bh}$--$M_{sph,*}$ relation existed.
However, we now know that the primary $M_{\rm bh}$--$M_{sph,*}$ relation is
not linear but quadratic-like.  Moreover, as noted by \citet{Angles:2013},
major mergers are not frequent enough to establish a linear relation in this
way.  The linear branch of the $M_{\rm bh}$--$M_{sph,*}$ relation, observed
only at the high-mass end of the $M_{\rm bh}$--$M_{sph,*}$ diagram, has
instead likely arisen from a few dry major mergers of galaxies having roughly the
same $M_{\rm bh}/M_{\rm sph,*}$ ratio \citep[e.g.][and references therein]{Dullo:2014}.

While hierarchical gas-rich merger models and AGN feedback models 
are highly valuable
\citep[e.g.][]{Fabian:1999,Wyithe:2003,Begelman:2005,Croton:2006,DiMatteo:2008,Natarajan:2012} 
the prediction (or use) of a linear $M_{\rm bh}$--$M_{\rm sph,*}$
relation does not describe the observed distribution for most galaxies. 
Indeed, any model which has predicted a 
linear $M_{\rm bh}$--$M_{\rm sph,*}$ relation, or that the 
black hole mass and the stellar mass of galactic spheroids should be 
proportional, or $\approx$0.001--0.002, for galaxies with 
$2\times10^6 \lesssim M_{\rm bh}/M_{\odot} \lesssim 2\times10^8$ 
$(M_{\rm sph} \approx 3-4\times10^{10} M_{\odot})$  
does not appear to match our Universe.  While black hole feedback
\citep[e.g.][]{Binney:1995,Ciotti:1997,Silk:1998}
likely regulates the black hole and host spheroid growth, the actual details 
are still a matter of debate.  Promisingly, 
\citet{Lu_Mo:2014} present a non-linear relation which, at stellar masses
above 10$^{10.5} M_{\odot}$, qualitatively matches the galaxy data from 
\citet{Scott:2013}. 

Evident in Figure~\ref{Fig2} is that the 35 galaxies 
from Table~\ref{Tab2} have higher spheroid masses relative to the
S\'ersic $M_{\rm bh}$--$M_{\rm sph,*}$ relation shown there.  As noted by
\citet{Busch:2014}, this might be due to an over-estimation of the stellar
mass-to-light ratios that they and we have used.  There is also room for
improvement in our $M_{\rm bh}$--$M_{\rm sph,*}$ relation for S\'ersic galaxies, and
Savorgnan et al.\ (in prep.) is working on deriving new spheroid masses from
Spitzer data for the S\'ersic (and core-S\'ersic) galaxies used here.  With
these advances, we will be in a better position to say if and how the AGN in
Figure~\ref{Fig2} are offset.  In passing we note two things.  First, using a
virial factor of 6 rather than 3 to determine the black hole masses would
(only) shift the AGN black hole masses by a factor of 2 higher in
Figure~\ref{Fig2}.  Second, elliptical galaxies and the bulges of disc
galaxies {\it may} follow offset sequences in the $M_{\rm bh}$--$M_{\rm
  sph,*}$ diagram, associated with the offset sequence in the $L$--$n$ diagram
discussed in \citet{Savorgnan:2013}.  Better quality data will however be
needed to test this issue.

Building on the Greene \& Ho (2007) sample of AGN candidates,
\citet{Dong:2012} report on 137 (309) AGN with virial black hole mass
estimates --- based on the the broad H$\alpha$ emission line --- ranging
from\footnote{One of the black hole mass estimates from \citet{Dong:2012} is
  $0.8\times10^5 M_{\odot}$.}  $2\times10^5$--$1.0\times10^6 M_{\odot}$
($2\times10^5$--$2\times10^6 M_{\odot}$).  They were able to refine their
sample from galaxies which may have broad H$\alpha$ lines due to star
formation processes rather than an AGN.  They showed that only about one-third
of their (typically Sbc galaxy) sample reside in the H{\sc II}, rather than
the Seyfert, region of diagnostic diagrams based on narrow-line ratios.
Future image decomposition should therefore provide yet more bulge
(i.e.\ spheroid) magnitudes and thus new data for the region of the $M_{\rm
  bh}$--$M_{\rm sph,*}$ diagram probed by the \citet{Jiang:2011a} data.

\subsection{Intermediate mass ($10^2$--$10^5 M_{\odot}$) black holes}

Using the steeper S\'ersic relation for
galaxies without depleted cores, intermediate mass black holes (IMBHs) have
now been predicted to exist in tens of galaxies possessing low-mass spheroids
and AGN activity \citep{GS:2013}.  Based on the original linear scaling
relation, the majority of these galaxies were thought to harbour black holes
with masses $\gtrsim$$10^6 M_{\odot}$ \citep{Dong:2006}.  Application of the
fundamental plane of black hole activity \citep{Merloni:2003,Falcke:2004} can
provide an independent estimate of their black hole mass, and we hope to apply
this technique to the above mentioned sample.  For example, \cite{GS:2013}
predict $\log(M_{\rm bh})=5.3\pm0.9$ for NGC~3185, while the black hole
fundamental plane predicts $\log(M_{\rm bh})=5.2\pm1.0$ (N.Webb et al.\ in
prep.).  Pushing to yet lower masses is obviously of importance to help
determine if and how the $M_{\rm bh}$--$M_{\rm sph,*}$ relation for S\'ersic
galaxies continues into the IMBH mass regime.
The implications of this near-quadratic relation are 
many, and while several were highlighted in our 2012--2013 papers, a couple
more specific examples are noted here.

First, the velocity dispersion of galaxy ASASSN-14ae suggests an IMBH of mass
$3\times10^4$ $M_{\odot}$ associated with the recent tidal disruption event
\citep{Holoien:2014}. The galaxy's young 2.2 Gyr age suggests it is a later
type spiral galaxy, and if we assign a bulge-to-total flux ratio of a tenth
\citep[e.g.,][]{GW:2008}, one obtains a bulge magnitude supportive of this
black hole mass when using the $M_{\rm bh}$--$M_{\rm sph,*}$ relation for
S\'ersic galaxies shown in Figure~\ref{Fig2}.  Over two dozen tidal
disruptions of stars by black holes are known \citep{Komossa:2013}, typically 
revealed via luminous X-ray flaring events.  We advocate obtaining accurate
bulge luminosities and masses for these galaxies because it will enable an 
additional estimate of the mass of the black holes, possibly supporting the
existence of yet more IMBH candidates.

Second, 
a recent simulation to predict the location of an intermediate mass black hole
from a spaghettified satellite galaxy around M31 may benefit from refined
initial conditions.  Given the initial satellite mass of $3\times10^9
M_{\odot}$ used by \citet{Miki:2014}, an $M_{\rm bh}$/$M_{\rm sph,*}$ mass
ratio of $4\times10^{-4}$ (derived from the $M_{\rm bh}$--$M_{\rm sph,*}$
relation for S\'ersic galaxies), rather than the (previously assumed constant)
value of $10^{-3}$ used by \citet{Miki:2014}, might be more appropriate.
Given the reduced amount of dynamical friction on a black hole which is 2.5
times less massive, it would be interesting to test how this alters their
suggested $0\degr.6 \times 0\degr.7$ search box for M31.  Observers may also
need to brace themselves for weaker observational signatures from such a
smaller black hole.

\citet{Jiang:2011a} revealed that their low-mass AGN-selected galaxies reside
below the near-linear $M_{\rm bh}$--$M_{\rm sph,*}$ relation (see also
Figure~\ref{Fig2}).  If large numbers of galaxies with $10^5 < M_{\rm
  bh}/M_{\odot} < 10^6$ had bulges that followed the near-linear $M_{\rm
  bh}$--$M_{\rm sph,*}$ relation, then they would have been found; that is,
there was no sample selection bias against AGN in small bulges.  However, a
small number of `bulgeless'\footnote{Care is required when identifying truly
  bulgeless galaxies from galaxies that simply have low (i.e.\ 5\%)
  bulge-to-disk flux ratios.}  galaxies containing AGN have been reported
\citep{Schramm:2013,Simmons:2013,Satyapal:2014}, and \citet{Jiang:2011b} noted
that 5\% of their sample had no bulges within the limits of their imaging
data.  While bulgeless galaxies reside on neither the near-linear nor the
near-quadratic $M_{\rm bh}$--$M_{\rm sph,*}$ relation --- because they have no
bulge --- they do serve to highlight that while the $M_{\rm bh}$--$M_{\rm
  sph,*}$ diagram has its main tracks, departures can exist. Moreover, if
these bulgeless galaxies were to undergo secular evolution of their disks and
form a pseudobulge, then at some point these pseudobulges would evolve
rightward in the $M_{\rm bh}$--$M_{\rm sph,*}$ diagram, possibly crossing, and
being found close to, the near-linear $M_{\rm bh}$--$M_{\rm sph,*}$ relation,
before presumably joining the majority of the low-mass bulges which appear to
define the near-quadratic $M_{\rm bh}$--$M_{\rm sph,*}$ relation.

Finally, we note that, 
based upon the hypothetical black hole mass estimates from \citet{Mieske:2013},
they reported that black holes in ultra-compact dwarf galaxies, and globular
clusters \citep[][but see \citet{Lanzoni:2014}]{Lutz:2013}, do not appear to
follow the near-linear $M_{\rm bh}$--$L_{\rm sph,*}$ relation defined by
galaxies with black holes predominantly more massive than $10^8 M_{\odot}$.
Given the high $M_{\rm bh}/M_{\rm UCD,*}$ ratios for their UCDs, and the
reported value of 15\% for M60-UCD1 \citep{Seth:2014}, UCDs are even more at
odds with the near-quadratic relation defined by low mass bulges in the $M_{\rm
  bh}$--$M_{\rm sph,*}$ diagram.  If UCDs are related to the stripped nuclei
of low mass galaxies, it may therefore be more appropriate to compare them
with relations pertaining to black holes in nuclear star clusters, for which
$M_{\rm bh}/M_{\rm nc}$ is already observed to reach $\sim$10\%
\citep{Graham:2009}.

\subsection{Nuclear star clusters}

There has been much attention on the relationship between black holes and
their host galaxies.  While large galaxies with depleted cores may be built
from dry merging events --- explaining their near-linear $M_{\rm bh}$--$M_{\rm
  sph,*}$ relation --- the S\'ersic galaxies and their black holes have formed
from more gaseous processes.  Most of these S\'ersic galaxies also contain a
nuclear star cluster \citep[e.g.][and references
  therein]{Baldass:2014,denBrok:2014}, which harbours and likely also feeds the
central massive black hole to some degree via stellar winds and also stellar
capture \citep[e.g.][and references therein]{Zhong:2014}.  Knowing the masses
of the black holes and also the nuclear star clusters is therefore of
interest.  

The high stellar density of nuclear star clusters may result in elevated
levels, relative to galaxies without nuclear star clusters, of inspiralling
stellar-mass black holes and neutron stars onto the central massive black
hole.  The physical size of these orbital decays, and thus the associated
orbital period, makes these `extreme mass ratio inspiral' 
\citep[EMRI:][and references therein]{Hils:1995,Rubbo:2006,Amaro:2014} 
events a likely source of gravitational 
radiation that could be detected by future space-based gravitational radiation
interferometers.  Of particular relevance here is that \citet{Mapelli:2012}
have shown how the reduced $M_{\rm bh}/M_{\rm sph,*}$ ratios --- from the
near-quadratic rather than near-linear $M_{\rm bh}$--$M_{\rm sph,*}$ relation
--- results in an order of magnitude lower number of such EMRI events, with
significant implications for the previously proposed Laser Interferometer
Space Antenna \citep[LISA:][]{LISA:1996}, currently replaced by the LISA
Pathfinder mission\footnote{\url{http://sci.esa.int/lisa-pathfinder/}} 
\citep[LPF:][]{Anza:2005,McNamara:2013} formerly known as SMART-2. 

The first attempt to quantify the coexistence of massive black holes 
(using directly measured black hole masses) and their 
surrounding nuclear star cluster can be found in 
\citet{Graham:2009}, with another recent work by \citet{Neumayer:2012},
but see also \citet{Gonz:2008} and \citet{Seth:2008}.  
One can gain further insight into their 
co-evolution by coupling the near-quadratic 
$M_{\rm bh}$--$M_{\rm sph,*}$ relation for S\'ersic galaxies 
with the $M_{\rm nc}$--$M_{\rm sph,*}$ relation \citep{GaG:2003,Balcells:2003}, 
where $M_{\rm nc}$ is the mass of the nuclear cluster of 
stars at the center of the galaxy.  This was first done in
\citet{Graham:2014a}, resulting in the 
discovery of the very steep $M_{\rm nc}$--$M_{\rm bh}$ relationship
as follows. 

\citet{SaG:2013} provide an updated $M_{\rm nc}$--$M_{\rm                           
  sph,*}$ relation in which $M_{\rm nc} \propto M_{\rm sph}^{0.55\pm0.15}$.
Most recently, a similar exponent of $0.57\pm0.05$ has been reported by 
\citet{denBrok:2014} based on $I$-band luminosities rather than masses.  
Therefore, using the approximation $M_{\rm bh} \propto M_{\rm sph,*}^2$
and $M_{\rm nc} \propto M_{\rm sph}^{0.6}$, one has $M_{\rm bh} \propto
M_{\rm nc}^{3.3}$.  
If, on the other hand, $M_{\rm nc} \propto M_{\rm sph}^{0.74}$ to $M_{\rm sph}^1$
\citep[e.g.][]{GaG:2003,Grant:2005,Cote:2006,Balcells:2007a}, 
one still has the rather steep relation $M_{\rm bh} \propto M_{\rm nc}^2$ to
$M_{\rm nc}^{2.7}$.  
That is, as one moves along the S\'ersic galaxy sequence, from 
$n<1$ in the low mass spheroids to values of $n$ around 4 in the higher mass
spheroids, the mass of the black hole is expected to grow dramatically faster than that of
the nuclear star cluster --- which is eventually eroded away in the core-S\'ersic
galaxies \citep{BekGra:2010}. 

This rapid growth was further checked in \citet{Graham:2014a} by combining the relation 
$M_{\rm bh}$--$\sigma^{5.5}$ \citep{Graham:2011} 
with the $M_{\rm nc}$--$\sigma^X$ relation.
Recent studies have reported values of $X$ equal to 
$1.57\pm0.24$ \citep{Graham:2012b}, 
$2.73\pm0.29$ \citep{Leigh:2012} and 
$2.11\pm0.31$ \citep{SaG:2013}.  Adopting a rough exponent of $X=2$, one has that 
$M_{\rm bh} \propto M_{\rm sph,*}^{2.77}$.  
It therefore seems apparent that the growth of black holes outstrips that of
nuclear star clusters, with a relation something like 
$M_{\rm bh} \propto M_{\rm sph,*}^{2.7\pm0.7}$ (given the current ranges presented above).

\subsection{Impact on next generation telescopes}

\citet{Do:2014} have predicted that the planned Thirty Meter Telescope
\citep[TMT][]{Sanders:2013} will be able to observe, i.e.\ resolve, the
sphere-of-influence of black holes in 100,000 galaxies.  They did so by
estimating the black hole masses using the {\it linear} $M_{\rm bh}$--$L_{\rm
  bulge}$ relation.  However, as first noted by \citet{Graham:2012a} and seen
in Figure~\ref{Fig2}, this relation 
dramatically over-estimates the black hole masses of low-luminosity bulges,
and will thus over-estimate the sphere-of-influence of the black holes
in such bulges.  This in turn results in an over-estimate to the numbers of
black holes whose sphere-of-influence will be resolved by the TMT.
It is beyond the
scope of this paper to re-derive the expected number, but this reduced number
should be of value to those developing the science objectives and
instrumentation for the billion dollar TMT \citep{Wright:2014,Moore:2014}, the
European Extremely Large Telescope \citep[E-ELT][]{Liske:2012,Evans:2014} and
the Giant Magellan Telescope \citep[GMT][]{Johns:2012,McGregor:2012}.

\subsection{Continued black hole growth in S\'ersic galaxies as a formation
  mechanism for over-massive black holes}

If a galaxy's gas reservoir is not expelled by feedback from its supermassive black
hole (perhaps due to accretion from a disk rather than isotropically), or is
replenished in a way that the supermassive black hole cannot prevent (perhaps
through a wet merger bringing in a significant amount of gas or through cold
accretion), then the supermassive black hole and its host spheroid will presumably
continue to follow the S\'ersic galaxy relation as gaseous processes will
still be dominant. Above spheroid stellar masses of $10^{11} M_\odot$, 
the S\'ersic relation is not well populated, therefore this
type of growth must be rare in massive galaxies, 
however there are now several potential examples
of this mechanism.

If the reported black hole mass in NGC~1277 is confirmed, this galaxy may be a
candidate for continued growth via gas-dominated processes beyond the
typical spheroid mass marked by the bend in the scaling relations. 
Other potential examples of this process include the ultra-massive black holes
hosted by brightest cluster galaxies (BCGs) identified by
\citet{Hlavacek:2012}. These black holes potentially became over-massive
relative to their host spheroids due to the widespread availability of gas
from the strong cooling flows found in the host clusters of the BCGs. Given
that the growth of these black holes is likely dominated by gaseous processes,
one might expect them to grow 
off and above the linear $M_{\rm bh}$--$M_{\rm sph,*}$ relation for core-S\'ersic
galaxies.

\subsection{Other explanations for outlying galaxies}
\label{sec:other}

A number of other physical processes can be responsible for changing the
location of a galaxy in the $M_{\rm bh}$--$M_{\rm sph,*}$ diagram. These
mechanisms either inhibit (or reverse) the growth of the stellar spheroid, or
somehow grow the black hole more rapidly than expected. Perhaps the most
commonly invoked mechanism for a galaxy appearing as an outlier in the left of
the $M_{\rm bh}$--$M_{\rm sph,*}$ diagram is tidal stripping of the
host galaxy. In this scenario a galaxy originally follows the black hole
scaling relations, but due to a tidal interaction with a nearby more-massive
neighbour it loses a significant fraction of its stellar mass, causing it to
move to the left in the $M_{\rm bh}$--$M_{\rm sph,*}$ diagram. This
scenario is discussed in more detail in \citet{Blom:2013}. The so-called
compact elliptical (cE) galaxy M32 is a well-known example of a galaxy thought
to be tidally stripped \citep{Dressler:1989,Bekki:2001,Graham:2002,Chilingarian:2009}
\citep[but see][]{Dierickx:2014}, and its position in the
$M_{\rm bh}$--$M_{\rm sph,*}$ diagram is consistent with this argument.
This type of galaxy is very rare, compared to the number of normal
galaxies, and their inclusion in the $M_{\rm bh}$--$M_{\rm sph,*}$ diagram
can therefore severely bias one's impression of what is happening at low
masses \citep[e.g.][their figure~8]{Graham:2007b,Sesana:2014}. 

In contrast, tidal stripping is unlikely to be
responsible for NGC~1277's position in the $M_{\rm bh}$--$M_{\rm sph,*}$
diagram because: i) it shows no signs of tidal disturbance
\citep{vandenBosch:2012}; ii) given its large mass, tidal stripping by a
companion would not be efficient in stripping away stars; and iii) to have
once been on the $M_{\rm bh}$--$M_{\rm sph,*}$, it would need to have
been stripped of an extreme $\sim 2 \times 10^{12}\ M_\odot$ of
stars, or $\sim$98\% of its spheroid mass. NGC~4486B is a less clear-cut
object, with a massive companion (M87) that could be responsible for the tidal
stripping of some of its stars. It is possible that both tidal stripping and
continued S\'ersic-mode growth played a role in NGC~4486B, with its black hole
growing unusually large through continued S\'ersic-mode growth, and then being
stripped of its stars through a tidal interaction.
Of course, the reported black masses in NGC~1277 and NGC~4486B may simply be
in error. This could happen due to velocity shear from an unresolved rotating
disk which elevates the central velocity dispersion, as noted in \citet{Graham:2011}.

Other possible pathways for producing outliers in the
$M_{\rm bh}$--$M_{\rm sph,*}$ diagram are less well explored. Mechanisms
for feeding gas onto a central black hole, while avoiding significant star
formation include: accretion from a dense, flat, gaseous disk or bar
\citep{Nayakshin:2012}, ultra-fast outflows \citep{Tombesi:2010} from AGN with
sub-Eddington accretion rates leading to enhanced black hole growth
\citep{Zubovas:2013} or the accretion of stellar in addition to gaseous
material, leading to optical or ultra-violet flaring events
\citep[e.g.][]{Komossa:2009,Gezari:2012,Donato:2014}. 
It is unclear whether
the effect of any of these mechanisms on a galaxy's position in the observed
$M_{\rm bh}$--$M_{\rm sph,*}$ relation would be significant, and further
theoretical work is required to better understand the potential affect of
these mechanisms and why they might preferentially exist in some galaxies ---
perhaps those with substantial central gas drag and nuclear star clusters or
discs.

Alternatively, a galaxy's gas supply may be removed by an external force
before its black hole grows large enough to remove that gas itself. This could
be accomplished by environmental effects such as ram pressure stripping
\citep{Gunn:1972} or the influence of a nearby AGN \citep{Shabala:2011}. Once
the gas is gone, the galaxy and black hole could, through dry mergers, grow
parallel to the dry merging / core-S\'ersic track in Figure \ref{Fig2},
populating the lower right part of the diagram.

\subsection{Summary}

We have collated data from seven different studies of low mass AGN.
We have unified their black hole mass derivations, and, when necessary, 
estimated the host spheroid's stellar mass. 
By combining this data, it is apparent in the $M_{\rm bh}$--$M_{\rm sph,*}$
diagram that it does not represent a collection of data pairs in which the 
black hole mass is randomly low, relative to the near-linear 
$M_{\rm bh}$--$M_{\rm sph,*}$ relation for core-S\'ersic galaxies.  
Such a scenario might be expected 
if the AGN reside in pseudobulges in which the black hole growth is lagging
that of the pseudobulge -- an idea floated in \citet{Hu:2008} and
\citet{Graham:2008a}.   
Simply because a galaxy resides below the near linear $M_{\rm
  bh}$--$M_{sph,*}$ relation does not imply that it is an offset pseudobulge.
Instead, the AGN with $M_{\rm bh} \gtrsim 10^6 M_{\odot}$ roughly 
overlap with the cloud of (predominantly inactive) S\'ersic galaxies which
define a near- or super-quadratic $M_{\rm bh}$--$M_{\rm sph,*}$ relation. 
They do however reside on the high $M_{\rm sph,*}$ edge of this cloud,
{\it possibly} indicating that their spheroid masses have been over-estimated
by us due to use of an overly large mass-to-light ratio conversion factor. 
The spheroids hosting AGN with $10^5 \lesssim M_{\rm bh}/M_{\odot} \lesssim 10^6$ closely track
this relation to lower masses. They also reveal that the (horizontal) scatter 
in the $\log(M_{\rm sph,*})$ direction is comparable at low and high masses. 
As such, the recent identification of separate scaling relations for 
S\'ersic and core-S\'ersic galaxies appears to explain many of the black holes
in AGN once thought to be under-massive.  

Through improvements in the quality of the $M_{\rm bh}$ and $M_{\rm sph,*}$
data, which are expected to be achievable in the near future via refined
virial factors, bulge/disc/bar decompositions of galaxy light, and stellar
mass-to-light ratios, we hope to further study the coevolution of black holes
with their host galaxy.  This will include determining if AGN may be offset
from the $M_{\rm bh}$--$M_{\rm sph,*}$ relation for S\'ersic galaxies defined
by largely inactive galaxies, and exploration into the realm of intermediate
mass black holes.

\acknowledgements
This research was supported by Australian Research Council funding through
grants DP110103509 and FT110100263.
This research made use of (i) the ``$K$-corrections calculator'' service available
at http://kcor.sai.msu.ru/ and (ii) the NASA/IPAC Extragalactic Database 
(NED: http://ned.ipac.caltech.edu).  

\bibliographystyle{apj}
\bibliography{}

\begin{thebibliography}{}

 \bibitem[Adelman-McCarthy et al.(2008)]{Adelman:2008} Adelman-McCarthy,
   J.~K., Ag{\"u}eros, M.~A., Allam, S.~S., et al.\ 2008, \apjs, 175, 297

 \bibitem[Agarwal et al.(2013)]{Agarwal:2013} Agarwal, B., Davis, A.~J.,
   Khochfar, S., Natarajan, P., \& Dunlop, J.~S.\ 2013, \mnras, 432, 3438

 \bibitem[Aguerri, Balcells \& Peletier (2001)]{Aguerri:2001}
Aguerri, J.A.L., Balcells, M., Peletier, R.F.\ 2001,
A\&A, 367, 428

 \bibitem[Alexander \& Hickox(2012)]{Alexander:2012} Alexander, D.~M., \&
   Hickox, R.~C.\ 2012, Nature, 56, 93

 \bibitem[Alonso-Herrero et al.(2013)]{Alonso:2013} Alonso-Herrero, A.,
   Pereira-Santaella, M., Rieke, G.~H., et al.\ 2013, \apj, 765, 78

 \bibitem[Amaro-Seoane et al.(2014)]{Amaro:2014} Amaro-Seoane, P., Gair, J.R.,
   Pound, A., Hughes, S.A., Sopuerta, C.F.\ 2014, Proceedings of the LISA
   Symposium X, Journal of Physics (arXiv:1410.0958)

 \bibitem[Angl{\'e}s-Alc{\'a}zar et al.(2013)]{Angles:2013}
   Angl{\'e}s-Alc{\'a}zar, D., {\"O}zel, F., \& Dav{\'e}, R.\ 2013, \apj, 770,
   5

\bibitem[Anza et al.(2005)]{Anza:2005} Anza, S., Armano, M., 
Balaguer, E., et al.\ 2005, Classical and Quantum Gravity, 22, 125 

 \bibitem[Assef et al.(2014)]{Assef:2014} Assef, R.J., Eisenhardt, R.M.,
   Stern, D., et al.\ 2014, ApJ, submitted (1408.1092)

\bibitem[Balcells et al.(2003)]{Balcells:2003} Balcells, M., Graham, A.~W.,
  Dom{\'{\i}}nguez-Palmero, L., \& Peletier, R.~F.\ 2003, \apjl, 582, L79

\bibitem[Balcells et al.(2007a)]{Balcells:2007a} Balcells, M., Graham, 
A.~W., \& Peletier, R.~F.\ 2007a, \apj, 665, 1084 

\bibitem[Balcells et al.(2007b)]{Balcells:2007b} Balcells, M., Graham,
A.~W., \& Peletier, R.~F.\ 2007b, \apj, 665, 1104

\bibitem[Baldassare et al.(2014)]{Baldass:2014}Baldassare, V.F., Gallo, E.,
  Miller, B.P., et al.\ 2014, ApJ, in press (arXiv:1406.6697)

\bibitem[Bardeen(1975)]{Bardeen:1975} Bardeen, J.~M.\ 1975, Dynamics 
of the Solar Systems, 69, 297 

\bibitem[Begelman \& Nath(2005)]{Begelman:2005} Begelman, M.~C., \& Nath,
  B.~B.\ 2005, \mnras, 361, 1387

 \bibitem[Beifiori et al.(2012)]{Beifiori:2012} Beifiori, A., 
Courteau, S., Corsini, E.~M., \& Zhu, Y.\ 2012, \mnras, 419, 2497 

 \bibitem[{{Bekki} {et~al.}(2001)}]{Bekki:2001} {Bekki}, k., {Couch} W.~J.,
   {Drinkwater}, M.~J., {Gregg}, M.~D., 2001, \apj, 557, 19

 \bibitem[Bekki (2010)]{Bekki:2010} Bekki, K.\ 2010, MNRAS, 401, L58

\bibitem[Bekki \& Graham(2010)]{BekGra:2010} Bekki, K., \& Graham,
  A.~W.\ 2010, \apjl, 714, L313

 \bibitem[Beletsky et al.(2011)]{Beletsky:2011} Beletsky, Y., Gadotti, D.~A.,
   Moiseev, A., Alves, J., \& Kniazev, A.\ 2011, \mnras, 418, L6

 \bibitem[Bell \& de Jong(2001)]{Bell:2001} Bell, E.~F., \& de Jong,
   R.~S.\ 2001, \apj, 550, 212

 \bibitem[Bellovary et al.(2014)]{Bellovary:2014} Bellovary, J.,
   Holley-Bockelmann, K., G{\"u}ltekin, K., et al.\ 2014, MNRAS, submitted
   (arXiv:1405.0286)



\bibitem[Benson et al.(2003)]{Benson:2003} Benson, A.~J., Bower, 
R.~G., Frenk, C.~S., et al.\ 2003, \apj, 599, 38 

\bibitem[Binney \& Tabor(1995)]{Binney:1995} Binney, J., \& Tabor, G.\ 1995,
  \mnras, 276, 663

 \bibitem[{{Blom} {et~al.}(2013)}]{Blom:2013} Blom, C., Forbes, D. A., Foster,
   C., Romanowsky, A. J., Brodie, J. P., 2013, MNRAS, in press (1401.5128)


 \bibitem[Bonoli et al.(2014)]{Bonoli:2014} Bonoli, S., Mayer, L., \&
   Callegari, S.\ 2014, \mnras, 437, 1576

 \bibitem[Booth \& Schaye(2009)]{Booth:2009} Booth, C.~M., \& Schaye,
  J.\ 2009, \mnras, 398, 53

 \bibitem[Bower et al.(2006)]{Bower:2006} Bower, R.~G., Benson, A.~J., Malbon,
   R., et al.\ 2006, \mnras, 370, 645

 \bibitem[{{Brown} {et~al.}(2013){Brown}, {Valluri}, {Shen}, \&
    {Debattista}}]{Brown:2013} {Brown}, J.~S., {Valluri}, M., {Shen}, J., \&
  {Debattista}, V.~P., 2013, \apj, 778, 151

 \bibitem[Busch et al.(2014)]{Busch:2014} Busch, G., Zuther, J., Valencia-S.,
  M., et al.\ 2014, \aap, 561, A140

 \bibitem[Cappellari et al.(2013)]{Cappellari:2013} Cappellari, M., 
McDermid, R.~M., Alatalo, K., et al.\ 2013, \mnras, 432, 1862 

 \bibitem[Chatzopoulos et al.(2014)]{Chatz:2014} Chatzopoulos, S., 
Fritz, T., Gerhard, O., et al.\ 2014, MNRAS, submitted, arXiv:1403.5266


 \bibitem[Chilingarian et al.(2008)]{Chilingarian:2008} Chilingarian, I.~V.,
   Cayatte, V., Durret, F., et al.\ 2008, \aap, 486, 85

 \bibitem[{{Chilingarian} {et~al.}(2009)}]{Chilingarian:2009}
{Chilingarian}, I., {Cayatte}, V., {Revas}, Y., {et~al.} 2009, Science, 326, 1379

 \bibitem[Chilingarian et al.(2010)]{Chilingarian:2010} Chilingarian, 
I.~V., Melchior, A.-L., \& Zolotukhin, I.~Y.\ 2010, \mnras, 405, 1409

\bibitem[Ciotti \& Ostriker(1997)]{Ciotti:1997} Ciotti, L., \& Ostriker,
  J.~P.\ 1997, \apjl, 487, L105

 \bibitem[Cody et al.(2009)]{Cody:2009} Cody, A.~M., Carter, D., 
Bridges, T.~J., Mobasher, B., \& Poggianti, B.~M.\ 2009, \mnras, 396, 1647 

 \bibitem[Collin et al.(2006)]{Collin:2006} Collin, S., Kawaguchi, T.,
  Peterson, B.~M., \& Vestergaard, M.\ 2006, \aap, 456, 75

\bibitem[Combes \& Sanders(1981)]{Combes:1981} Combes, F., \& Sanders,
  R.~H.\ 1981, \aap, 96, 164

 \bibitem[Conroy et al.\ (2010)]{Conroy:2014}Conroy, C., van Dokkum, P., 
Kravtsov, A., ApJL, submitted (arXiv:1406.3026)

\bibitem[C{\^o}t{\'e} et al.(2006)]{Cote:2006} C{\^o}t{\'e}, P., 
Piatek, S., Ferrarese, L., et al.\ 2006, \apjs, 165, 57 

 \bibitem[Cox(2000)]{Cox:2000} Cox, A.~N.\ 2000, Allen's 
Astrophysical Quantities, 4th ed. Publisher: New York: AIP Press; Springer,
2000. Edited by Arthur N. Cox

 \bibitem[{{Croton} {et~al.}(2006){Croton}, {Springel}, {White}, {De Lucia},
  {Frenk}, {Gao}, {Jenkins}, {Kauffmann}, {Navarro}, \&
  {Yoshida}}]{Croton:2006}
{Croton}, D.~J., {Springel}, V., {White}, S.~D.~M., {et~al.} 2006, \mnras, 365,
  11

\bibitem[Danzmann \& LISA Study(1996)]{LISA:1996} Danzmann, K., et al.\
  1996, Classical and Quantum Gravity, 13, 247


 \bibitem[Davies et al.(1983)]{Davies:1983} Davies, R.~L., 
Efstathiou, G., Fall, S.~M., Illingworth, G., 
\& Schechter, P.~L.\ 1983, \apj, 266, 41 

 \bibitem[Debattista et al.(2013)]{Debattista:2013} Debattista, V.~P., 
Kazantzidis, S., \& van den Bosch, F.~C.\ 2013, \apj, 765, 23 

 \bibitem[Del Moro et al.(2013)]{Del:2013} Del Moro, A., 
Alexander, D.~M., Mullaney, J.~R., et al.\ 2013, \memsai, 84, 665 

\bibitem[den Brok et al.(2013)]{denBrok:2014} den Brok, M., Peletier, R.F.,
  Seth, A., et al.\ 2014, MNRAS, in press

 \bibitem[de Rijcke et 
al.(2005)]{deRijcke:2005} de Rijcke, S., Michielsen, D., Dejonghe, H.,
   Zeilinger, W.~W., \& Hau, G.~K.~T.\ 2005, \aap, 438, 491 

\bibitem[Diamond-Stanic \& Rieke(2012)]{2012ApJ...746..168D} Diamond-Stanic,
  A.~M., \& Rieke, G.~H.\ 2012, \apj, 746, 168

 \bibitem[Dierickx et al.(2014)]{Dierickx:2014} Dierickx, M., Blecha, 
L., \& Loeb, A.\ 2014, \apjl, 788, L38

\bibitem[Di Matteo et al.(2008)]{DiMatteo:2008} Di Matteo, T., Colberg, J.,
  Springel, V., Hernquist, L., \& Sijacki, D.\ 2008, \apj, 676, 33


 \bibitem[Do et al.(2014)]{Do:2014} Do, T., Wright, S.~A., 
Barth, A.~J., et al.\ 2014, AJ, 147, 93

 \bibitem[Dominguez-Tenreiro, Tissera \& Saiz (1998)]{Dom:1998}
Dominguez-Tenreiro, R., Tissera, P.B., Saiz, A.\ 1998, Ap\&SS, 263, 35

 \bibitem[{{Donato} {et~al.}(2013){Donato}, {Cenko}, {Covino},
     {et~al.}}]{Donato:2014} {Donato}, D., {Cenko}, S.P., {Covino}, S.,
   {et~al.} 2014,\apj, 781, 59


 \bibitem[Dong et al.(2012)]{Dong:2012} Dong, X.-B., Ho, L.~C., 
Yuan, W., et al.\ 2012, \apj, 755, 167

 \bibitem[Dong \& De Robertis(2006)]{Dong:2006} Dong, X.~Y., \& De Robertis, M.~M.\
2006, \aj, 131, 1236 

 \bibitem[D'Onofrio(2001)]{D'Onofrio:2001} D'Onofrio, M.\ 2001, \mnras, 326,
  1517

\bibitem[Dressler(1989)]{Dressler:1989} Dressler, A.\ 1989, Active 
Galactic Nuclei, 134, 217

 \bibitem[Driver et al.(2008)]{Driver:2008} Driver, S.~P., Popescu, 
C.~C., Tuffs, R.~J., et al.\ 2008, \apjl, 678, L101

 \bibitem[Drouart et al.(2014)]{Drouart:2014}Drouart, G., De Breuck, C.,
  Vernet, J. et al.\ 2014, A\&A, 566, A53

 \bibitem[Dubois et al.(2012)]{Dubois:2012} Dubois, Y., Devriendt, 
J., Slyz, A., \& Teyssier, R.\ 2012, \mnras, 420, 2662 

 \bibitem[Dullo \& Graham(2013)]{Dullo:2013} Dullo, B.~T., \& Graham, A.~W.\
  2013, ApJ, 768, 36

 \bibitem[Dullo \& Graham(2014)]{Dullo:2014} Dullo, B.~T., \& Graham, A.~W.\
  2014, MNRAS, 444, 2700

\bibitem[Emsellem et al.(2007)]{SAURON:2007} Emsellem, E., 
Cappellari, M., Krajnovi{\'c}, D., et al.\ 2007, \mnras, 379, 401 

 \bibitem[Emsellem et al.(2011)]{Emsellem:2011} Emsellem, E., 
Cappellari, M., Krajnovi{\'c}, D., et al.\ 2011, \mnras, 414, 888 


 \bibitem[Evans et al.(2014)]{Evans:2014} Evans, C.~J., Puech, M., 
Barbuy, B., et al.\ 2014, arXiv:1406.6369 

 \bibitem[Faber et al.(1997)]{Faber:1997} Faber, S.~M., Tremaine, 
S., Ajhar, E.~A., et al.\ 1997, \aj, 114, 1771 

\bibitem[Fabian(1999)]{Fabian:1999} Fabian, A.~C.\ 1999, \mnras, 
308, L39 

 \bibitem[Falcke et al.(2004)]{Falcke:2004} Falcke, H., K{\"o}rding,
  E., \& Markoff, S.\ 2004, \aap, 414, 895

 \bibitem[{{Fanidakis} {et~al.}(2013){Fanidakis}, {Georgakakis},
  {Mountrichas}, {Krumpe}, {Baugh}, {Lacey}, {Frenk}, {Miyaji}, \&
  {Benson}}]{Fanidakis:2013}
{Fanidakis}, N., {Georgakakis}, A., {Mountrichas}, G., {et~al.} 2013, \mnras,
submitted (arXiv:1305.2200) 

 \bibitem[Feng, Shen \& Li (2014)]{Feng:2014} Feng, H., Shen, Y., Li,
   H.\ 2014, ApJ, in press (arXiv:1408.6952)




 \bibitem[Ferrarese \& Merritt(2000)]{FaM:2000} Ferrarese, L., \& Merritt, D.\ 2000, \apjl, 539, L9 

\bibitem[Forbes et al.(2008)]{Forbes:2008} Forbes, D.~A., Lasky, 
P., Graham, A.~W., \& Spitler, L.\ 2008, \mnras, 389, 1924

 \bibitem[{{Gabor} \& {Bournaud}(2013)}]{Gabor:2013}
{Gabor}, J.~M., \& {Bournaud}, F. 2013, \mnras, in press (arXiv:1306.2954)


 \bibitem[Gebhardt et al.(2000)]{Geb:2000} Gebhardt, K., Bender, 
R., Bower, G., et al.\ 2000, \apjl, 539, L13 

 \bibitem[Gezari et al.(2012)]{Gezari:2012}
{Gezari}, S., {Chornock}, R., {Rest}, A., {et~al.} 2012, \nat, 485, 217

\bibitem[Gonz{\'a}lez Delgado et al.(2008)]{Gonz:2008} Gonz{\'a}lez
  Delgado, R.~M., P{\'e}rez, E., Cid Fernandes, R., \& Schmitt, H.\ 2008, \aj,
  135, 747

 \bibitem[{{Graham}(2002)}]{Graham:2002} Graham, A.~W. 2002, \apj, 568, 13

 \bibitem[Graham(2007a)]{Graham:2007a} ---. 2007a, Bulletin of 
the American Astronomical Society, 39, 759 

\bibitem[Graham(2007b)]{Graham:2007b} ---. 2007b, \mnras, 379, 711 

 \bibitem[{{Graham}(2008a)}]{Graham:2008a} ---. 2008a, \apj, 680, 143

 \bibitem[Graham(2008b)]{Graham:2008b} ---. 2008b, PASA, 25, 167

 \bibitem[{{Graham}(2012a)}]{Graham:2012a} ---. 2012a, \apj, 746, 113

 \bibitem[{{Graham}(2012b)}]{Graham:2012b} ---. 2012b, \mnras, 422, 1586


 \bibitem[{{Graham}(2013)}]{Graham:2013} ---. 2013, in ``Planets, Stars and Stellar Systems'', Vol. 6, p.91-140, T.D.Oswalt \& W.C Keel (eds.), Springer Publishing (arXiv:1108.0997)

 \bibitem[{{Graham}(2014a)}]{Graham:2014a} ---. 2014a, in Structure and
   Dynamics of Disk Galaxies, Edited by M.S.~Seigar and P.~Treuthardt. ASP
   Conference Series, 480, 185

 \bibitem[{{Graham}(2014b)}]{Graham:2014b} ---. 2014b, in Star Clusters and
   Black Holes in Galaxies Across Cosmic Time, IAU Symp.\ 312, submitted

\bibitem[Graham \& Driver(2005)]{2005PASA...22..118G} Graham, A.~W., \&
  Driver, S.~P.\ 2005, PASA, 22, 118

\bibitem[Graham et al.(2003)]{Graham:2003} Graham, A.~W., Erwin, 
P., Trujillo, I., \& Asensio Ramos, A.\ 2003, AJ, 125, 2951

\bibitem[Graham \& Guzm{\'a}n(2003)]{GaG:2003} Graham, A.~W., \&
  Guzm{\'a}n, R.\ 2003, \aj, 125, 2936

\bibitem[Graham \& Li(2009)]{GandLi:2009} Graham, A.~W., \& Li,
  I.-h.\ 2009, \apj, 698, 812

 \bibitem[Graham et al.(2011)]{Graham:2011} Graham, A.~W., Onken, 
C.~A., Athanassoula, E., \& Combes, F.\ 2011, \mnras, 412, 2211 (arXiv:1007.3834)

 \bibitem[{{Graham} \& {Scott}(2013)}]{GS:2013}
{Graham}, A.~W., \& {Scott}, N. 2013, \apj, 764, 151

 \bibitem[Graham et al.(2014)]{Korea:2014} Graham, A.~W., Scott, N.,
   Schombert, J.~M.\ 2014, Pub.\ Korean Astronomical Society, submitted

 \bibitem[{{Graham} \& {Spitler}(2009)}]{Graham:2009}
{Graham}, A.~W., \& {Spitler}, L.~R. 2009, \mnras, 397, 2148

 \bibitem[Graham \& Worley(2008)]{GW:2008} Graham, A.~W., \& Worley,
  C.~C.\ 2008, \mnras, 388, 1708

\bibitem[Grant et al.(2005)]{Grant:2005} Grant, N.~I., Kuipers, 
J.~A., \& Phillipps, S.\ 2005, \mnras, 363, 1019 


 \bibitem[Greene \& Ho(2007)]{2007ApJ...670...92G} Greene, J.~E., \& Ho,
   L.~C.\ 2007, \apj, 670, 92

 \bibitem[Greene et al.(2010)]{Greene:2010} Greene, J.~E., Peng, 
C.~Y., Kim, M., et al.\ 2010, \apj, 721, 26 

 \bibitem[G{\"u}ltekin et al.(2009)]{Gultekin:2009} G{\"u}ltekin, K., 
Richstone, D.~O., Gebhardt, K., et al.\ 2009, \apj, 698, 198

 \bibitem[{{Gunn} \& {Gott}(1972)}]{Gunn:1972}
{Gunn}, J.~E., \& {Gott}, III, J.~R. 1972, \apj, 176, 1

 \bibitem[Haehnelt et al.(1998)]{Haehnelt:1998} Haehnelt, M.~G., 
Natarajan, P., \& Rees, M.~J.\ 1998, \mnras, 300, 817

 \bibitem[{{H{\"a}ring} \& {Rix}(2004)}]{Haring:2004}
{H{\"a}ring}, N., \& {Rix}, H.-W. 2004, \apjl, 604, L89

 \bibitem[{{Hartmann} {et~al.}(2013){Hartmann}, {Debattista}, {Cole},
   {Valluri}, {Widrow}, \& {Shen}}]{Hartmann:2013}
{Hartmann}, M., {Debattista}, V.P., {Cole}, D.R., {Valluri}, M., 
 {Widrow}, L.M., {Shen}, J. 2013, \mnras, submitted (arXiv:1309.2634)

 \bibitem[Held et al.(1992)]{Held:1992} Held, E.~V., de Zeeuw, T., 
Mould, J., \& Picard, A.\ 1992, \aj, 103, 851 

\bibitem[Hils \& Bender(1995)]{Hils:1995} Hils, D., \& Bender, P.~L.\ 1995,
  \apjl, 445, L7

 \bibitem[Hirschmann et al.(2010)]{Hirschmann:2010} Hirschmann, M., 
Khochfar, S., Burkert, A., et al.\ 2010, \mnras, 407, 1016 

 \bibitem[{{Hlavacek-Larrondo} {et~al.}(2012){Hlavacek-Larrondo}, {Fabian},
    {Edge}, \& {Hogan}}]{Hlavacek:2012} {Hlavacek-Larrondo}, J., {Fabian},
  A.~C., {Edge}, A.~C., \& {Hogan}, M.~T. 2012, \mnras, 424, 224

 \bibitem[Ho(1999)]{Ho:1999} Ho, L.\ 1999, Observational 
Evidence for the Black Holes in the Universe, 234, 157 

\bibitem[Hohl(1975)]{Hohl:1975} Hohl, F.\ 1975, Dynamics of the 
Solar Systems, 69, 349 

\bibitem[Hohl \& Zang(1979)]{Hohl:1979} Hohl, F., \& Zang, T.~A.\ 1979, \aj,
  84, 585

 \bibitem[Holoien et al.(2014)]{Holoien:2014} Holoien, T.~W.-S., Prieto, J.~L.,
  Bersier, D., et al.\ 2014, MNRAS, submitted (arXiv:1405.1417)


 \bibitem[{{Hu}(2008)}]{Hu:2008}{Hu}, J. 2008, \mnras, 386, 2242

 \bibitem[Jahnke \& Macci{\`o}(2011)]{Jahnke:2011} Jahnke, K., \&
  Macci{\`o}, A.~V.\ 2011, \apj, 734, 92

\bibitem[Jerjen \& Binggeli(1997)]{JaB:1997} Jerjen, H., \& Binggeli, B.\ 1997, The
  Nature of Elliptical Galaxies; 2nd Stromlo Symposium, 116, 239 

 \bibitem[Jiang et al.(2011a)]{Jiang:2011a} Jiang, Y.-F., Greene, 
J.~E., \& Ho, L.~C.\ 2011a, \apjl, 737, L45 

 \bibitem[Jiang et al.(2011b)]{Jiang:2011b} Jiang, Y.-F., Greene, 
J.~E., Ho, L.~C., Xiao, T., \& Barth, A.~J.\ 2011b, \apj, 742, 68 

 \bibitem[{{Jiang} {et~al.}(2013){Jiang}, {Ho}, {Dong}, {Yang}, \&
  {Wang}}]{Jiang:2013}
{Jiang}, N., {Ho}, L.~C., {Dong}, X.-B., {Yang}, H., \& {Wang}, J. 2013, \apj,
  770, 3

 \bibitem[Johns et al.(2012)]{Johns:2012} Johns, M., McCarthy, P., 
Raybould, K., et al.\ 2012, \procspie, 8444, 

 \bibitem[Karouzos et al.(2013)]{Karouzos:2014} Karouzos, M., Im, M., 
Trichas, M., et al.\ 2014, ApJ, submitted arXiv:1309.7353

 \bibitem[Kaspi et al.(2000)]{Kaspi:2000} Kaspi, S., Smith, P.~S., 
Netzer, H., et al.\ 2000, \apj, 533, 631 

 \bibitem[Kawata \& Gibson(2005)]{Kawata:2005} Kawata, D., \& Gibson, B.~K.\
  2005, \mnras, 358, L16

 \bibitem[Khandai et al.(2012)]{Khandai:2012} Khandai, N., Feng, Y., 
DeGraf, C., Di Matteo, T., \& Croft, R.~A.~C.\ 2012, \mnras, 423, 2397

 \bibitem[Khandai et al.(2014)]{Khandai:2014} Khandai, N., Di Matteo, T.,
  Croft, R., et al. 2014, \mnras, submitted (arXiv:1402.0888)

 \bibitem[{{Komossa} {et~al.}(2009){Komossa}, {Zhou}, {Rau}, {Dopita},
  {Gal-Yam}, {Greiner}, {Zuther}, {Salvato}, {Xu}, {Lu}, {Saxton}, \&
  {Ajello}}]{Komossa:2009}
{Komossa}, S., {Zhou}, H., {Rau}, A., {et~al.} 2009, \apj, 701, 105

\bibitem[Komossa(2013)]{Komossa:2013} Komossa, S.\ 2013, IAU 
Symposium, 290, 53 

\bibitem[Kormendy(1982)]{Kormendy:1982} Kormendy, J.\ 1982, Saas-Fee 
Advanced Course 12: Morphology and Dynamics of Galaxies, 113 

\bibitem[Kormendy(1993)]{Kormendy:1993} Kormendy, J.\ 1993, Galactic Bulges,
  IAU Symp., 153, 209

\bibitem[Kormendy \& Bender(2011)]{Kormendy:2011} Kormendy, J., \& Bender, R.\ 2011, \nat,
  469, 377 

\bibitem[Kormendy \& Kennicutt(2004)]{KK:2004} Kormendy, J., \& Kennicutt, R.~C.,
  Jr.\ 2004, \araa, 42, 603 

\bibitem[Kormendy \& Richstone(1995)]{1995ARA&A..33..581K} Kormendy, J., \&
  Richstone, D.\ 1995, ARA\&A, 33, 581




 \bibitem[Kourkchi et al.(2012)]{Kourkchi:2012} Kourkchi, E., 
Khosroshahi, H.~G., Carter, D., et al.\ 2012, \mnras, 420, 2819

 \bibitem[LaMassa et al.(2013)]{LaMassa:2013} LaMassa, S.~M., 
Heckman, T.~M., Ptak, A., \& Urry, C.~M.\ 2013, \apjl, 765, L33

\bibitem[Lanzoni(2014)]{Lanzoni:2014} Lanzoni, B.\ in Star Clusters and
   Black Holes in Galaxies Across Cosmic Time, IAU Symp.\ 312

\bibitem[Laor(1998)]{1998ApJ...505L..83L} Laor, A.\ 1998, \apj, 505, L83

 \bibitem[Laor(2000)]{Laor:2000} Laor, A.\ 2000, \apjl, 543, L111 

 \bibitem[Laor(2001)]{Laor:2001} Laor, A.\ 2001, \apj, 553, 677 


\bibitem[Leigh et al.(2012)]{Leigh:2012} Leigh, N., B{\"o}ker, T., 
\& Knigge, C.\ 2012, \mnras, 424, 2130

 \bibitem[Lehmer et al.(2013)]{Lehmer:2013} Lehmer, B.~D., Lucy, 
A.~B., Alexander, D.~M., et al.\ 2013, \apj, 765, 87

 \bibitem[Licquia \& Newman(2014)]{Licquia:2014} Licquia, T.C., \& Newman, J.A.\
  2014, ApJ, submitted (arXiv:1407.1078)

 \bibitem[Liske et al.(2012)]{Liske:2012} Liske, J., Padovani, P., 
\& Kissler-Patig, M.\ 2012, \procspie, 8444, 

 \bibitem[Lu \& Mo (2014)]{Lu_Mo:2014} Lu, Z., Mo, H.J.\ 2014, ApJL, submitted (arXiv:1407.4382)

 \bibitem[L{\"u}tzgendorf et al.(2013)]{Lutz:2013} L{\"u}tzgendorf, N.,
  Kissler-Patig, M., Neumayer, N., et al.\ 2013, \aap, 555, A26

 \bibitem[{{Magorrian} {et~al.}(1998){Magorrian}, {Tremaine}, {Richstone},
  {Bender}, {Bower}, {Dressler}, {Faber}, {Gebhardt}, {Green}, {Grillmair},
  {Kormendy}, \& {Lauer}}]{Magorrian:1998}
{Magorrian}, J., {Tremaine}, S., {Richstone}, D., {et~al.} 1998, \aj, 115, 2285

\bibitem[Mapelli et al.(2012)]{Mapelli:2012} Mapelli, M., Ripamonti,
  E., Vecchio, A., Graham, A.~W., \& Gualandris, A.\ 2012, \aap, 542, A102

\bibitem[McGaugh \& Schombert(2014)]{MaS:2014} McGaugh, S.S., Schombert,
  J.M.\ 2014, AJ, 148, 77

  \bibitem[Matkovi{\'c} \& Guzm{\'a}n(2005)]{Matkovic:2005} Matkovi{\'c}, A., \& Guzm{\'a}n, R.\
2005, \mnras, 362, 289 

 \bibitem[{{Marconi} \& {Hunt}(2003)}]{Marconi:2003}
{Marconi}, A., \& {Hunt}, L.~K. 2003, \apjl, 589, L21

 \bibitem[{{Mathur} {et~al.}(2012){Mathur}, {Fields}, {Peterson}, \&
    {Grupe}}]{Mathur:2012} {Mathur}, S., {Fields}, D., {Peterson}, B.~M., \&
  {Grupe}, D. 2012, \apj, 754, 146

\bibitem[McConnell \& Ma(2013)]{McConnell:2013} McConnell, N.~J., \& Ma,
  C.-P.\ 2013, \apj, 764, 184

 \bibitem[McConnell et al.(2011)]{McConnell:2011} McConnell, N.~J., Ma, C.-P.,
  Gebhardt, K., et al.\ 2011, \nat, 480, 215 (arXiv:1112.1078)

 \bibitem[McGregor et al.(2012)]{McGregor:2012} McGregor, P.~J., 
Bloxham, G.~J., Boz, R., et al.\ 2012, \procspie, 8446, 

\bibitem[McLure \& Dunlop(2001)]{2001MNRAS.327..199M} McLure, R.~J., \&
  Dunlop, J.~S.\ 2001, \mnras, 327, 199

\bibitem[McNamara(2013)]{McNamara:2013} McNamara, P.~W.\ 2013, International
  Journal of Modern Physics D, 22, 41001

 \bibitem[Merloni \& Heinz(2008)]{Merloni:2008} Merloni, A., \& Heinz,
  S.\ 2008, \mnras, 388, 1011

 \bibitem[Merloni et al.(2003)]{Merloni:2003} Merloni, A., Heinz, S., \& di
   Matteo, T.\ 2003, \mnras, 345, 1057

 \bibitem[Merritt \& Ferrarese(2001a)]{MaF:2001a} Merritt, D., \&
   Ferrarese, L.\ 2001a, \apj, 547, 140

\bibitem[Merritt \& Ferrarese(2001b)]{MaF:2001b} Merritt, D., \&
  Ferrarese, L.\ 2001b, \mnras, 320, L30

 \bibitem[Mieske et al.(2013)]{Mieske:2013} Mieske, S., Frank, M.~J.,
  Baumgardt, H., et al.\ 2013, \aap, 558, A14

 \bibitem[Miki et al.(2014)]{Miki:2014} Miki, Y., Mori, M., Kawaguchi, T.,
  Saito, Y.\ 2014, \apj, in press (arXiv:1401.4645)

 \bibitem[Monari et al.(2014)]{Monari:2014} Monari, G., Antoja, T., \& Helmi,
   A.\ 2014, (arXiv:1306.2632)

 \bibitem[Moore et al.(2014)]{Moore:2014} Moore, A.~M., Larkin, J.~E., Wright,
   S.~A., et al.\ 2014, arXiv:1407.2995

\bibitem[Natarajan \& Volonteri(2012)]{Natarajan:2012} Natarajan, P., \&
  Volonteri, M.\ 2012, \mnras, 422, 2051

 \bibitem[{{Nayakshin} {et~al.}(2012){Nayakshin}, {Power}, \&
  {King}}]{Nayakshin:2012}
{Nayakshin}, S., {Power}, C., \& {King}, A.~R. 2012, \apj, 753, 15

\bibitem[Neumayer \& Walcher(2012)]{Neumayer:2012} Neumayer, N., \&
  Walcher, C.~J.\ 2012, Advances in Astronomy, 2012,

 \bibitem[Onken et al.(2004)]{Onken:2004} Onken, C.~A., Ferrarese, 
L., Merritt, D., et al.\ 2004, \apj, 615, 645 

 \bibitem[Page et al.(2012)]{Page:2012} Page, M.~J., Symeonidis, 
M., Vieira, J.~D., et al.\ 2012, \nat, 485, 213 

 \bibitem[Park et al.(2012)]{Park:2012} Park, D., Kelly, B.~C., 
Woo, J.-H., \& Treu, T.\ 2012, \apjs, 203, 6 

 \bibitem[Peng(2007)]{Peng:2007} Peng, C.~Y.\ 2007, \apj, 671, 1098

 \bibitem[Peterson \& Wandel(2000)]{PaW:2000} Peterson, B.~M., \& Wandel, A.\ 2000, \apjl, 540, L13 

 \bibitem[Poggianti(1997)]{Poggianti:1997} Poggianti, B.~M.\ 1997, \aaps,
  122, 399 

\bibitem[Querejeta et al.(2014)]{Querejeta:2014} Querejeta, M., Eliche-Moral,
  M.C., Tapia, T., Borlaff, A., Rodriguez-P\'erez, C.,  Zamorano, J., Gallego,
  J.\ 2014, A\&A, in press (arXiv:1409.5126)

 \bibitem[Reines et al.(2013)]{Reines:2013} Reines, A.~E., Greene, 
J.~E., \& Geha, M.\ 2013, \apj, 775, 116

\bibitem[Rubbo et al.(2006)]{Rubbo:2006} Rubbo, L.~J., 
Holley-Bockelmann, K., \& Finn, L.~S.\ 2006, \apjl, 649, L25 

 \bibitem[Rusli et al.(2011)]{Rusli:2011} Rusli, S.~P., Thomas, J., 
Erwin, P., et al.\ 2011, \mnras, 410, 1223 

 \bibitem[Saha, Martinez-Valpuesta \& Gerhard (2012)]{Saha:2012} Saha K.,
   Martinez-Valpuesta I., Gerhard O. 2012, MNRAS, 421, 333

 \bibitem[Sanders(2013)]{Sanders:2013} Sanders, G.~H.\ 2013, Journal of
  Astrophysics and Astronomy, 34, 81

 \bibitem[Sanghvi et al.(2014)]{Sanghvi:2014} Sanghvi, J., Kotilainen, J.,
   Falomo, R., Decarli, R., Karhunen, K., Uslenghi, M.\ 2014, \mnras, in press
   (arXiv:1409.1948)

 \bibitem[Sani et al.(2011)]{Sani:2011} Sani, E., Marconi, A., 
Hunt, L.~K., \& Risaliti, G.\ 2011, \mnras, 413, 1479 

\bibitem[Satyapal et al.(2014)]{Satyapal:2014} Satyapal, S., Secrest, 
 N.~J., McAlpine, W., et al.\ 2014, \apj, 784, 113

 \bibitem[Savorgnan et al.(2013)]{Savorgnan:2013} Savorgnan, G., 
Graham, A.~W., Marconi, A., et al.\ 2013, \mnras, 434, 387 

 \bibitem[Schlafly \& Finkbeiner(2011)]{Schlafly:2011} Schlafly, E.~F., \&
  Finkbeiner, D.~P.\ 2011, \apj, 737, 103 


\bibitem[Schombert \& Smith(2012)]{Schombert:2012} Schombert, J., \&
  Smith, A.~K.\ 2012, PASA, 29, 174

\bibitem[Schramm et al.(2013)]{Schramm:2013} Schramm, M., Silverman, 
 J.~D., Greene, J.~E., et al.\ 2013, \apj, 773, 150 

\bibitem[Scott \& Graham(2013)]{SaG:2013} Scott, N., \& Graham,
  A.~W.\ 2013, \apj, 763, 76

 \bibitem[{{Scott} {et~al.}(2013){Scott}, {Graham}, \& {Schombert}}]{Scott:2013}
{Scott}, N., {Graham}, A.~W., \& {Schombert}, J. 2013, \apj, 768, 76

\bibitem[Sesana et al.(2014)]{Sesana:2014} Sesana, A., Barausse, E., Dotti,
  M., Rossi, E.M.\ 2014, ApJ, 794, 104

\bibitem[Seth et al.(2008)]{Seth:2008} Seth, A., Ag{\"u}eros, M.,
  Lee, D., \& Basu-Zych, A.\ 2008, \apj, 678, 116

\bibitem[Seth et al.(2014)]{Seth:2014} Seth, A., van den Bosch, R., Mieske,
  S., et al.\ 2014, Nature, 513, 398

 \bibitem[Seymour et al.(2012)]{Seymour:2012} Seymour, N., Altieri, 
B., De Breuck, C., et al.\ 2012, \apj, 755, 146 

 \bibitem[{{Shabala} {et~al.}(2011){Shabala}, {Kaviraj}, \&
  {Silk}}]{Shabala:2011}
{Shabala}, S.~S., {Kaviraj}, S., \& {Silk}, J. 2011, \mnras, 413, 2815

\bibitem[Sijacki et al.(2014)]{Sijacki:2014} Sijacki, D., Vogelsberger, M.,
  Genel, S., et al.\ 2014, MNRAS, submitted (arXiv:1408.6842)

 \bibitem[Silk \& Rees(1998)]{Silk:1998} Silk, J., \& Rees, M.~J.\ 1998, \aap,
  331, L1

\bibitem[Simmons et al.(2013)]{Simmons:2013} Simmons, B.~D., 
 Lintott, C., Schawinski, K., et al.\ 2013, \mnras, 429, 2199

\bibitem[Skrutskie et al.(2006)]{2MASS:2006} Skrutskie, M.~F., 
Cutri, R.~M., Stiening, R., et al.\ 2006, \aj, 131, 1163 

 \bibitem[Springel et al.(2005)]{Springel:2005} Springel, V., Di 
Matteo, T., \& Hernquist, L.\ 2005, \mnras, 361, 776 

 \bibitem[Straatman et al.(2014)]{Straatman:2014} Straatman, C.~M.~S., 
Labb{\'e}, I., Spitler, L.~R., et al.\ 2014, ApJL, submitted arXiv:1312.4952 

 \bibitem[Taylor et al.(2011)]{Taylor:2011} Taylor, E.~N., Hopkins, 
A.~M., Baldry, I.~K., et al.\ 2011, \mnras, 418, 1587 

 \bibitem[Thornton et al.(2008)]{Thornton:2008} Thornton, C.~E., Barth, A.~J.,
  Ho, L.~C., Rutledge, R.~E., \& Greene, J.~E.\ 2008, \apj, 686, 892

 \bibitem[{{Tombesi} {et~al.}(2010){Tombesi}, {Cappi}, {Reeves}, {Palumbo},
  {Yaqoob}, {Braito}, \& {Dadina}}]{Tombesi:2010}
{Tombesi}, F., {Cappi}, M., {Reeves}, J.~N., {et~al.} 2010, \aap, 521, A57

 \bibitem[Tortora et al.(2009)]{Tortora:2009} Tortora, C., 
Napolitano, N.~R., Romanowsky, A.~J., Capaccioli, M., 
\& Covone, G.\ 2009, \mnras, 396, 1132

 \bibitem[Trakhtenbrot \& Netzer(2012)]{Trakhtenbrot:2012} Trakhtenbrot, B., \&
  Netzer, H.\ 2012, \mnras, 427, 3081

 \bibitem[Trujillo et al.(2014)]{Trujillo:2014} Trujillo, I., 
Ferr{\'e}-Mateu, A., Balcells, M., Vazdekis, A., 
\& S{\'a}nchez-Bl{\'a}zquez, P.\ 2014, \apjl, 780, L20 

 \bibitem[{{van den Bosch} {et~al.}(2012){van den Bosch}, {Gebhardt},
  {G{\"u}ltekin}, {van de Ven}, {van der Wel}, \& {Walsh}}]{vandenBosch:2012}
{van den Bosch}, R.~C.~E., {Gebhardt}, K., {G{\"u}ltekin}, K., {et~al.} 2012,
  \nat, 491, 729

\bibitem[Vika et al.(2012)]{Vika:2012} Vika, M., Driver, S.~P., 
Cameron, E., Kelvin, L., \& Robotham, A.\ 2012, \mnras, 419, 2264

 \bibitem[Walsh et al.(2012)]{Walsh:2012} Walsh, J.~L., van den 
Bosch, R.~C.~E., Barth, A.~J., \& Sarzi, M.\ 2012, \apj, 753, 79 

\bibitem[Wandel(1999)]{1999ApJ...519L..39W} Wandel, A.\ 1999, \apjl, 519, L39

\bibitem[Webster et al.(1995)]{Webster:1995} Webster, R.~L., Francis, P.~J.,
  Peterson, B.~A., Drinkwater, M.~J., \& Masci, F.~J.\ 1995, \nat, 375, 469


\bibitem[Worthey(1994)]{Worthey:1994} Worthey, G.\ 1994, \apjs, 95, 107 

 \bibitem[Wright et al.(2014)]{Wright:2014} Wright, S.~A., Larkin, 
J.~E., Moore, A.~M., et al.\ 2014, arXiv:1407.2996

 \bibitem[{{Wurster} \& {Thacker}(2013)}]{Wurster:2013}
{Wurster}, J., \& {Thacker}, R.J. 2013, \mnras, 431, 2513

\bibitem[Wyithe \& Loeb(2003)]{Wyithe:2003} Wyithe, J.~S.~B., \& Loeb,
  A.\ 2003, \apj, 595, 614

\bibitem[Wyse et al.(1997)]{Wyse:1997} Wyse, R.~F.~G., Gilmore, G., \& Franx,
  M.\ 1997, \araa, 35, 637 

\bibitem[Yee(1992)]{1992ASPC...31..417Y} Yee, H.~K.~C.\ 1992, in Relationships
  Between Active Galactic Nuclei and Starburst Galaxies, ed. A. V. Filippenko,
  ASP Conference Series (ASP: San Francisco), 31, 417


 \bibitem[Yuan et al.(2014)]{Yuan:2014} Yuan, W., Zhou, H., Dou, L., Dong,
  X.-B., Fan, X., Wang, T.-G.\ 2014, \apj, 782, 55

\bibitem[Zhong et al.(2014)]{Zhong:2014} Zhong, S., Berczik, P., Spurzem,
  R.\ 2014, ApJ, 792, 137

 \bibitem[{{Zubovas} \& {King}(2013)}]{Zubovas:2013}
{Zubovas}, K., \& {King}, A. 2013, \apj, 769, 51

\end{thebibliography}

\end{document}